\documentclass[english,journal=jctcce,manuscript=article,email=true,etalmode=truncate,maxauthors=0,doi=true]{achemso}

\usepackage{soul}
\usepackage{graphicx}
\graphicspath{{figures/}}
\usepackage{physics}
\usepackage{multirow}
\usepackage{xcolor}
\usepackage{caption}
\usepackage{subcaption}
\usepackage{bm}
\usepackage{amsfonts}
\usepackage{yfonts} 
\usepackage[cal=boondoxo,bb=stix]{mathalpha}  
\usepackage[title,titletoc]{appendix}
\usepackage{diagbox}
\usepackage{booktabs} 
\usepackage{threeparttable}
\usepackage{float}

\usepackage{xspace} 

\usepackage{longtable}
\usepackage{algorithm}
\usepackage{algpseudocode}

\usepackage{hyperref} 
\hypersetup{colorlinks=true, citecolor=blue, urlcolor=blue, linkcolor=blue}
\usepackage{cleveref}
  	\crefname{figure}{Figure}{Figures}
  	\crefname{table}{Table}{Tables}
  	\crefname{equation}{Eq.}{Eqs.}
  	\crefname{section}{Section}{Sections}
  	\crefname{subsection}{Section}{Sections}
  	\crefname{subsubsection}{Section}{Sections}
  	\crefname{algorithm}{Algorithm}{Algorithms}
    \crefname{appendix}{Appendix}{Appendices}
\newcommand{\doi}[1]{\href{http://dx.doi.org/#1}{\nolinkurl{#1}}}

\newcommand{\code}[1]{\texttt{#1}}
\newcommand\vartextvisiblespace[1][.5em]{%
  \makebox[#1]{%
    \kern.07em
    \vrule height.3ex
    \hrulefill
    \vrule height.3ex
    \kern.07em
  }
}

\usepackage{todonotes}

\newcommand{\name}{MADF\xspace}

\title{Physics-Driven Construction of Compact Primitive Gaussian Density Fitting Basis Sets} 

\author{Kshitijkumar A. Surjuse}
\author{Edward F. Valeev}
\affiliation{Department of Chemistry, Virginia Tech, Blacksburg, VA 24061}
\email{efv@vt.edu}

\begin{document}

\date{\today}

\newpage

\begin{abstract}
We present model-assisted density fitting (MADF) basis set generator, an algorithm for generating primitive atomic Gaussian density fitting (DF) basis sets (DFBSs) from a contracted Gaussian orbital basis set (OBS). The \name algorithm produces DFBSs suitable for accurate robust DF approximation of 2-particle interactions in mean-field and correlated electronic structure. The algorithm is designed to (a) saturate the OBS product space by a large regularized set of primitive solid-harmonic Gaussian shells with nonuniform distribution of exponents followed by (b) pruning of the shells according to their contributions to the 2-body energy of a correlated atomic ensemble. Building the DFBS generator model almost exclusively on mathematical and physical principles allows one to limit the number of parameters that control the density fitting error to three, with a single set of parameters sufficient for computations with all basis cardinal numbers, with and without correlation of core electrons, with and without scalar and spin-dependent relativistic effects, spanning almost all of the Periodic Table. Performance assessment included basis sets up to quadruple-zeta quality from several major basis set families, using molecules composed of main-group, d-block, and f-block elements. The resulting DF errors in Hartree-Fock and second-order MP2 energies (with relativistic all-electron treatments, when appropriate) were on the order of $20$ and $10$ $\mu E_h$ per electron, respectively.
\end{abstract}

\maketitle 

\section{Introduction}\label{sec:intro}

Density fitting (DF)\cite{VRG:whitten:1973:JCP, VRG:baerends:1973:CP, VRG:mintmire:1982:PRA, VRG:dunlap:2000:JMST,  KAS:dunlap:2010:MP,VRG:hu:2013:CJC}, also known as resolution of identity (RI),\cite{KAS:vahtras:1993:CPL, VRG:kendall:1997:TCA} is a widely used cost and complexity reduction technique for approximate evaluation of various operators in the electronic structure theory, most commonly the electron repulsion integrals (ERIs).
The cost of ERI evaluation and their storage are often crucial limitations in practical simulations of electronic structure. The factorization of ERIs using the DF approximation can be exploited in various ways to gain significant speedups and reduce storage requirements\cite{VRG:feyereisen:1993:CPL, KAS:eichkorn:1995:CPL, VRG:manby:2003:JCP, KAS:aquilante:2007:JCPa, KAS:kelley:2013:JCP,  VRG:hollman:2014:JCP, KAS:delcey:2015:JCP, KAS:reynolds:2018:JCP}, such as to reduce the cost of electrostatic potential evaluation\cite{KAS:eichkorn:1995:CPL, KAS:weigend:2002:PCCP} and to facilitate correlated and relativistic computations\cite{KAS:weigend:2002:PCCP,VRG:deprinceiii:2011:JCTC, VRG:nagy:2016:JCTC, VRG:peng:2019:IJQC, KAS:bintrim:2022:JCTC}. 
DF has also recently emerged as the route to deeper factorizations of the ERI tensor such as real-space tensor hypercontraction,\cite{VRG:hohenstein:2012:JCP} algebraic pseudospectral,\cite{VRG:pierce:2021:JCTC} canonical polyadic,\cite{VRG:pierce:2023:JCTC} and interpolative separable density fitting (ISDF)\cite{KAS:lu:2015:JCP}.
Although our sole focus in this work is on the Gaussian AOs,
note that density fitting is a key enabling technology for other choices of AOs such as Slater-type\cite{KAS:tevelde:2001:JCC} and numerical AOs.\cite{KAS:ren:2012:NJP}

In the DF approach, the AO products (or ``densities") in the ERIs are expanded in a special-purpose density-fitting (or auxiliary) basis of AOs.
Traditionally, density-fitting basis sets (DFBSs) are manually-optimized for each orbital basis set (OBS) in ad hoc manner\cite{KAS:weigend:1998:CPL, VRG:weigend:2002:JCP, KAS:weigend:2002:PCCP, KAS:hattig:2005:PCCP, KAS:weigend:2006:PCCPa,  KAS:weigend:2008:JCC, VRG:yousaf:2008:JCP, KAS:hattig:2012:PCCP, KAS:hill:2013:IJQC, KAS:tanaka:2013:JCC, KAS:hellweg:2015:PCCP}. 
For efficiency reasons not only are DFBS matched to the OBS but they are also matched to specific use cases; e.g., DFBS for Coulomb fitting (RI-J)\cite{VRG:weigend:2002:JCP,KAS:weigend:2006:PCCPa}, Coulomb and exchange fitting (RI-JK)\cite{KAS:weigend:2002:PCCP, KAS:weigend:2008:JCC} or for the second-order M{\o}ller-Plesset (MP2) correlation energy (RI-C)\cite{KAS:weigend:1998:CPL, KAS:weigend:2002:PCCP, KAS:hattig:2005:PCCP, VRG:yousaf:2008:JCP, KAS:hattig:2012:PCCP, KAS:tanaka:2013:JCC, KAS:hellweg:2015:PCCP}.
DFBS optimization is usually done with cumbersome iterative HF and MP2 calculations on a training set of atoms and molecules.

Due to the significant effort involved in the DFBS development, there are only a few OBS families for which there are matching DFBS, and those often only focus on the top half of the Periodic Table. Among the commonly used basis sets, only the correlation consistent basis sets and the def2- (TURBOMOLE) basis sets have matching DFBS.
Even for these families there are signficant coverage gaps; e.g., correlation-consistent OBSs such as cc-pVXZ, cc-pCVXZ, and cc-pwCVXZ\cite{VRG:dunning:1989:JCP,VRG:dunning:2001:JCP,KAS:peterson:2002:JCP} have DFBS coverage gaps in the first to third rows of the Periodic Table.\cite{VRG:pritchard:2019:JCIM}
Only the def2- family has a thorough DFBS coverage for atoms up to Rn (with atoms heavier than Kr using effective core potential).\cite{KAS:weigend:2002:PCCP, KAS:weigend:2006:PCCPa, VRG:pritchard:2019:JCIM,KAS:franzke:2023:JCTC, KAS:weigend:2003:JCP, KAS:weigend:2005:PCCP} Unfortunately, for heavier elements,
there are no useful DFBS coverage.
For example, OBS families such as ANO-RCC basis sets\cite{KAS:roos:2004:TCA, KAS:roos:2005:JPCA,  KAS:roos:2008:JPCA}, cc-pwCVXZ-DK / -DK3 / -X2C\cite{KAS:balabanov:2005:JCP, VRG:lu:2016:JCP, VRG:peterson:2015:JCP, VRG:feng:2017:JCP} basis sets, the x2c-XZVP\cite{VRG:pollak:2017:JCTC, KAS:franzke:2020:JCTC} family of basis sets, and Dyall basis sets\cite{VRG:gomes:2010:TCA, KAS:dyall:2012:TCA, KAS:dyall:2023:} do not have the corresponding DFBS designed for correlation calculations. 
The only two relativistic DFBSs available in Basis set exchange (BSE)\cite{VRG:pritchard:2019:JCIM} database are x2c-JFIT and x2c-universal-JFIT\cite{KAS:franzke:2020:JCTC}, but they are designed only for Coulomb fitting and are not suitable for many-body correlated simulations or even for hybrid Kohn-Sham DFT computations.

To address the challenges of ad hoc optimization of DFBS many black-box methods have been proposed for {\em generating} (rather than optimizing) DFBS for a given OBS have been proposed\cite{VRG:yang:2007:JCPa,KAS:hellmann:2022:JCTC,KAS:diaz-tinoco:2025:JCTC,KAS:antonini:2025:JPCA,KAS:stoychev:2017:JCTC,KAS:lehtola:2021:JCTC, KAS:lehtola:2023:JCTC}
. A DFBS generator can be viewed as an OBS $\to$ DFBS map controlled by zero or more model parameters. The key difference between DFBS generators and manual optimization is that no nonlinear optimization is involved in the definition of the map. A map can be defined straightforwardly to produce exact atomic DFBS (= DFBS that is exact for any computation on a single atom), but such DFBS is too large to be practical. Nevertheless, such an {\em exact} map is typically the starting point for the design of approximate DFBS generators. Nearly all DFBS generators in practice are approximate, i.e. they produce DFBS that are not exact even for a single atom.

Most DBFS generators are designed only for DF in the context of evaluating the Coulomb potential of a charge density.
These include the AutoABS method by Yang et al.\cite{VRG:yang:2007:JCPa}, atomic Cholesky decomposition (aCD) and atomic compact Cholesky decomposition (acCD)\cite{KAS:aquilante:2007:JCP, KAS:aquilante:2009:JCP}, long-range corrected auxiliary basis sets by Hellmann et al.\cite{KAS:hellmann:2022:JCTC}, even tempered auxiliary basis sets with shared exponents by Daz-Tinoco et al.\cite{KAS:diaz-tinoco:2025:JCTC} and a similar approach has also recently been reported for relativistic OBSs\cite{KAS:antonini:2025:JPCA}. 

 DFBS generators designed for mean-field and correlated electronic structure simulation include the AutoAux procedure by Stoychev et al.,\cite{KAS:stoychev:2017:JCTC} methods proposed by
Lehtola\cite{KAS:lehtola:2021:JCTC, KAS:lehtola:2023:JCTC}, and \code{PySCF} built-in generator of even-tempered DFBSs.

The AutoAux method, available in the \code{ORCA} software package\cite{VRG:neese:2020:JCP} and in the command-line interface of Basis Set Exchange (BSE)\cite{VRG:pritchard:2019:JCIM}, produces a primitive even-tempered DFBS that can be applied universally (in mean-field and correlated simulations).
The AutoAux generator uses even-tempered sequences of Gaussian primitive AOs to span the OBS-deduced range of exponents for each angular momentum block. The exponent ranges and even-tempered ratios vary with the angular momenta ($L$) and the atomic number $Z$ (these even-tempering parameters are prescribed for up to $L=7$; the BSE implementation of AutoAux uses the same exponent ratio for $L\geq 7$). The generator also limits the highest angular momentum of DFBS AOs below what is needed for exact atomic DF, in a manner that depends on the highest angular momentum of the occupied orbitals. AutoAux as a result is quite complex, defined with more than a dozen model parameters in total.

\code{PySCF}'s simple built-in generator\cite{pu2025enhancingpyscfbasedquantumchemistry} produces even-tempered primitive DFBS and is very similar in spirit to AutoAux. It has fewer model parameters (e.g., even-tempering ratio is not graded with angular momenta) and claims to offer similar accuracy to AutoAux, although no extensive comparison exists.

Lehtola presented a DFBS generator that produces nearly exact atomic DFBS composed of primitive solid-harmonic Gaussian (SHG) AOs.\cite{KAS:lehtola:2021:JCTC}
First, the OBS AOs on a given atom are used to generate a base pool of primitive SHGs by approximating each angular momentum channel of solid-harmonic OBS AO products by a single solid-harmonic AO. The resulting pool is then regularized by pivoted Cholesky decomposition (pCD) of the two-center two-electron Coulomb integral matrix, with
the accuracy of the resulting primitive DFBS controlled robustly by a single threshold. Although the resulting DFBSs are nearly exact, they are extremely large, lead to high condition numbers in molecules, and contain primitive Gaussians with twice the angular momentum of the highest angular momentum ($L_{\text{OBS}}$) of functions in the parent OBS. 
Hellmann et al.\cite{KAS:hellmann:2023:JPCA} extended Lehtola's method\cite{KAS:lehtola:2021:JCTC} to include exact spherical functions with mixed angular momentum functions instead of creating multiple angular momentum channels; however, it does not address the large number of primitives in the generated DFBS.
Lehtola addressed the latter issue in his recent work\cite{VRG:lehtola:2023:JCTC} by extending the algorithm to prune high-angular momentum functions out of the DFBS with a tunable parameter.
The algorithm then uses Kállay's\cite{KAS:kallay:2014:JCP} contraction method with singular value decomposition (SVD) to contract the DFBS, further reducing its size.

An attractive alternative to density fitting that avoids these issues and provides some of its advantages is the (pivoted) Cholesky decomposition (CD).\cite{VRG:beebe:1977:IJQC}
Any positive-definite 2-particle interaction represented in an arbitrary OBS can be decomposed with accuracy controlled by a single threshold.
CD has been used for electronic structure simulations in both relativistic\cite{KAS:banerjee:2023:JCP, KAS:zhang:2024:JCTC,KAS:chamoli:2025:JCTC} and nonrelativistic\cite{KAS:koch:2003:JCP, KAS:pedersen:2009:TCA, KAS:aquilante:2011:LTiCCaP} contexts and close connections between CD and DF have been demonstrated\cite{KAS:pedersen:2024:WCMS}.
Although the accuracy of CD can be robustly pushed beyond the reach of conventional density fitting, the cost of
molecular CD can be substantially higher than that of DF-based approaches and CD is not as universally applicable black-box technology as desired.\cite{KAS:krisiloff:2015:MaCM} 
The use of CD for construction of atomic density fitting basis sets, under the name of atomic CD (aCD), was pioneered by Aquilante et al.\cite{KAS:aquilante:2007:JCP} The improved version of aCD, termed atomic compact CD (acCD),\cite{ VRG:aquilante:2009:JCP} was proposed later that greatly reduced the number of primitive and contracted DF AOs.
Both the aCD and acCD approaches can be viewed as black-box 1-parameter DFBS generators and are also unbiased towards mean-field and correlated methods, albeit they produce deeply contracted DFBS AOs.

To navigate the severe gaps in the coverage by existing manually-constructed DFBS and the notable shortcomings of the existing DFBS generators we attempted to design a new DFBS generator that produces (a) {\em primitive} Gaussian DFBSs, (b)
of comparable size and accuracy as the manually-optimized DFBS, (c) usable in the context of mean-field and correlated simulation, and
(d) has as few adjustable parameters as possible to ensure maximum universality. By combining the existing ideas for spanning the product space of OBS AOs with new ideas for regularizing and pruning the exact DFBS using atomic correlated ensemble RDMs we arrived at a DFBS generator dubbed \name. The purpose of this manuscript is to describe it and assess its performance.
Following a quick recap of density fitting in \cref{sec:df} we describe
the 2 key ingredients of the \name generator: the regularization of the complete set of candidate primitives (described in \cref{sec:products-of-gaussians}) and subsequent pruning by a 2-body energy estimator model (described in \cref{sec:estimating-2bE-DF-contribution}).
\cref{sec:technical} describes the technical and details of the parameter training regiment.
In \cref{sec:results} we document the performance of the DFBSs generated by our method against the manually-optimized DFBSs and the DFBSs generated by the AutoAux algorithm and acCD. 

\section{Formalism}\label{sec:formalism}
\subsection{Density Fitting} \label{sec:df}
DF approximates product of AOs (``density'') $\phi_\mu({\bf r}) \phi_\nu({\bf r}) \equiv ({\bf r}|\mu\nu)$ 
as a linear combination of DFBS AOs $\phi_X({\bf r})$
\begin{equation}\label{eq:df-approximation}
    |\mu\nu) \approx \sum_X C_{\mu\nu}^X |X) \equiv |\widetilde{\mu\nu}).
\end{equation}
The fitting coefficients $C_{\mu\nu}^X$ are obtained by minimizing a norm of the density error $|\delta_{\mu\nu}) \equiv |\mu\nu) - |\widetilde{\mu\nu})$. To minimize the error in
the diagonal elements of the matrix representation of a positive operator $\hat{O}$, namely in $(\mu\nu|\hat{O}|\mu\nu)$, the optimal choice of the error norm is the expectation value of operator $\hat{O}$ itself (as is done throughout this work):
\begin{align}
|| f ||_{\hat{O}} \equiv (f|\hat{O}|f).
\end{align}
Such choice makes the error in $(\mu\nu|\hat{O}|\rho\sigma)$ quadratic in the density fitting errors and is a special case of {\em robust} density fitting;\cite{VRG:dunlap:1979:JCP, KAS:dunlap:2009:IJQC}
if the fitting norm were defined with operator $\hat{W} \neq \hat{O}$ the explicit robust fitting approach would be necessary and variational extensions\cite{VRG:reine:2008:JCP} could be introduced for convenience.
Least-squares minimization of $||\delta_{\mu \nu}||_{\hat{O}}$ is equivalent to solving the following equation
\begin{align}
\label{eq:rDF}
    \hat{O}|\mu \nu) = \sum_Y \hat{O}|Y) C^Y_{\mu\nu}.
\end{align}
In {\em global} DF \cref{eq:rDF} is projected on every DF AO in the system to produce the usual system defining the fitting coefficients:
\begin{align}
\label{eq:grDF}
    (X|\hat{O}|\mu \nu) = \sum_Y (X|\hat{O}|Y) C^Y_{\mu\nu}.
\end{align}

Clearly, DF is a misnomer (and so is RI), as in the most common scenario, what is being ``fitted'' (minimized) is the $L^2$ norm of $\hat{O}^{1/2}|\delta_{\mu \nu})$, not of $|\delta_{\mu \nu})$ itself.
For the most important case where $\hat{O}$ is the Coulomb potential operator $\hat{J}$,
\begin{align}
    ({\bf r}|\hat{J}|f) \equiv \int \mathrm{d} \mathbf{r}' \, |{\bf r} - {\bf r}'|^{-1} f({\bf r'}),
\end{align}
the robust fitting minimizes the $L^2$ norm of the fitting error in the {\em electric field} generated by the AO product.

DFBS generators usually\cite{VRG:yang:2007:JCPa,KAS:hellmann:2022:JCTC,KAS:diaz-tinoco:2025:JCTC,KAS:antonini:2025:JPCA,KAS:stoychev:2017:JCTC,KAS:lehtola:2021:JCTC,KAS:lehtola:2023:JCTC,VRG:neese:2020:JCP} work in a single-atom context, i.e., DFBS is constructed using information about products on OBS AOs on a single atom. It is reasonable to question whether this is sufficient. Assuming a system composed of a single atom type, global density fitting in a system of $n$ atoms will be able to fit every such 1-center product of OBS AOs using $n$ times as many DFBS AOs as in a single atom. Indeed, for long-range $\hat{O}$ the fitting of even strongly-localized AO products generally involves DFBS AOs from far away.\cite{VRG:gill:2005:JCP} Thus, 1-center products in a molecule can borrow DFBS AOs from other centers to make the fitting more accurate than in a single atom. This suggests that a single atom poses a more challenging DF setting than a molecule, which supports the use of a single-atom setting for the generation of DFBS.

\subsection{Products of primitive Gaussians AOs}\label{sec:products-of-gaussians}

The density fitting becomes exact if the span of DFBS matches the span of the OBS AO products. Because DFBS AOs are atom-centered but the products of AOs in a molecule are generally centered between atoms, exact DF is only possible for a single atom. We then confine ourselves to a single atom. Furthermore, we will limit ourselves to the case of OBS and DFBS composed of solid-harmonic Gaussian (SHG) AOs. Unnormalized primitive SHG with exponent $\alpha$ is defined
as:\cite{VRG:helgaker:2000:,VRG:rico:2012:IJQC}
\begin{align}
    \chi_{\alpha, l, m}({\bf r}) = & r^l \exp(-\alpha r) y_{lm}(\theta, \phi) \\
     y_{lm}(\theta, \phi) = & (-1)^m P_l^{|m|}(\theta) \begin{cases}\cos(m \phi), \quad m \geq 0 \\
     \sin(|m|\phi), \quad m<0
     \end{cases}.
\end{align}
Whereas product of any number of concentric primitive Cartesian Gaussians is a single Cartesian Gaussian (modulo normalization), product of concentric SHGs $\chi_{\alpha_\mu, l_\mu, m_\mu}$ and $\chi_{\alpha_\nu, l_\nu, m_\nu}$ is a linear combination of $r^{l_\mu + l_\nu - L} \chi_{\alpha_\mu+\alpha_\nu, L, m_\mu + m_\nu}$ for $\max(|l_\mu - l_\nu|,m_\mu + m_\nu) \leq L \leq l_\mu + l_\nu$. Clearly, the contribution to the product from the Clebsch-Gordan channel $L < l_{\mu} + l_{\nu}$ is an SHG multiplied by $r^{l_\mu + l_\nu - L}$. Following Stoychev\cite{KAS:stoychev:2017:JCTC}, Lehtola proposed to account for this effect by approximating each such contribution by a single SHG of angular momentum $L$ with the following effective exponent (see Appendix II of Ref. \citenum{KAS:lehtola:2021:JCTC}):
\begin{equation}\label{eq:eff-exponent}
    \alpha_\text{eff}^{L} = \left[ \frac{\Gamma(l_{\mu} +l_{\nu} + \frac{3}{2})\Gamma(L+2)}{\Gamma(L+ \frac{3}{2})\Gamma(l_{\mu} +l_\nu + 2)} \right]^2 (\alpha_{\mu} + \alpha_{\nu})
    \begin{cases}
    = \alpha_\mu+\alpha_\nu, \quad L = l_\mu+l_\nu \\
    > \alpha_\mu+\alpha_\nu, \quad L < l_\mu+l_\nu
    \end{cases}.
\end{equation}
We follow this prescription here to define the ``complete'' set of candidate primitive SHGs for a given product of primitive SHG OBS AOs.

\begin{figure}
\centering
\begin{subfigure}{.5\textwidth}
  \centering
  \includegraphics[width=1.0\linewidth]{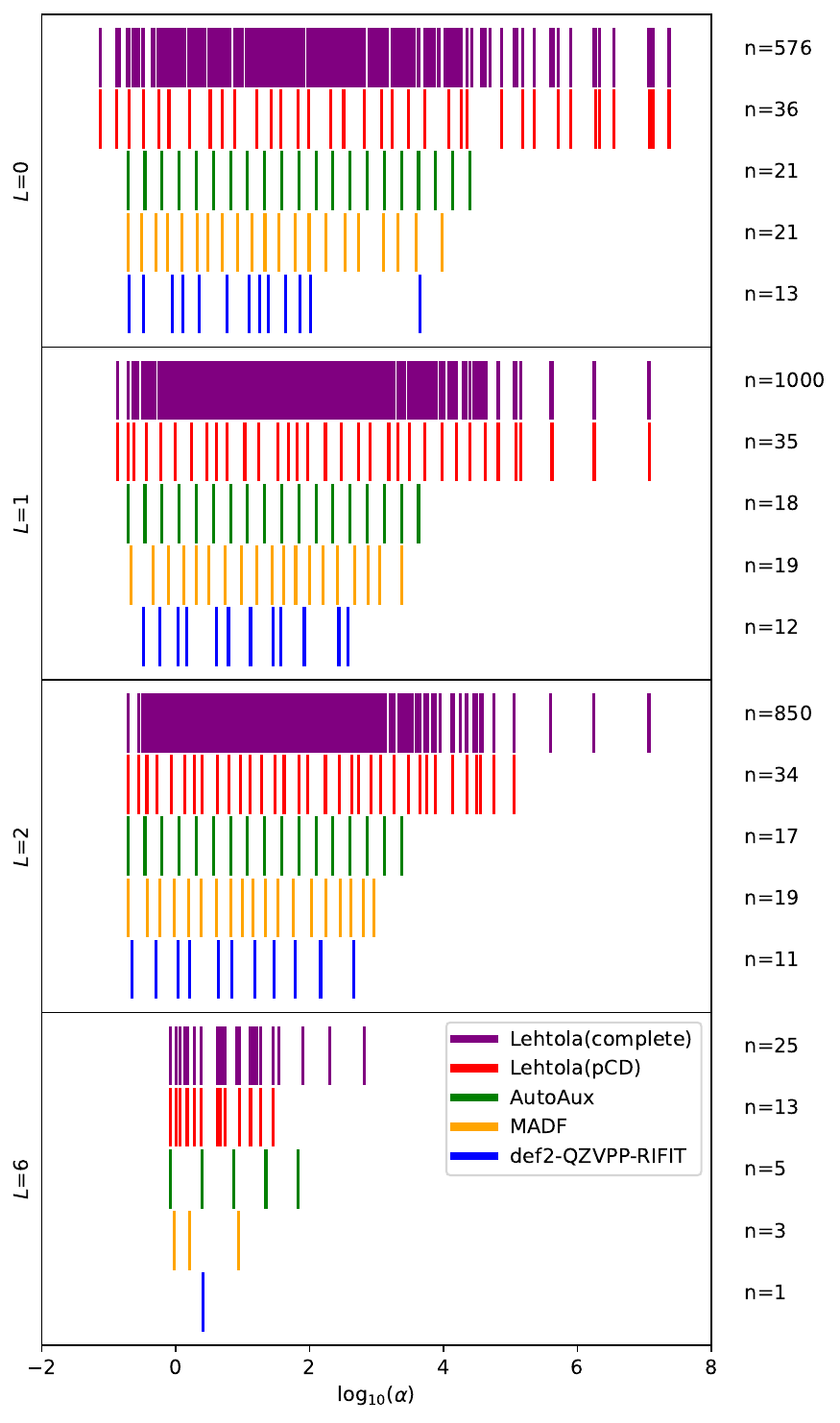}
  \caption{OBS = def2-QZVPP}
  \label{fig:nonrel-exponent-distribution}
\end{subfigure}%
\begin{subfigure}{.5\textwidth}
  \centering
  \includegraphics[width=1.0\linewidth]{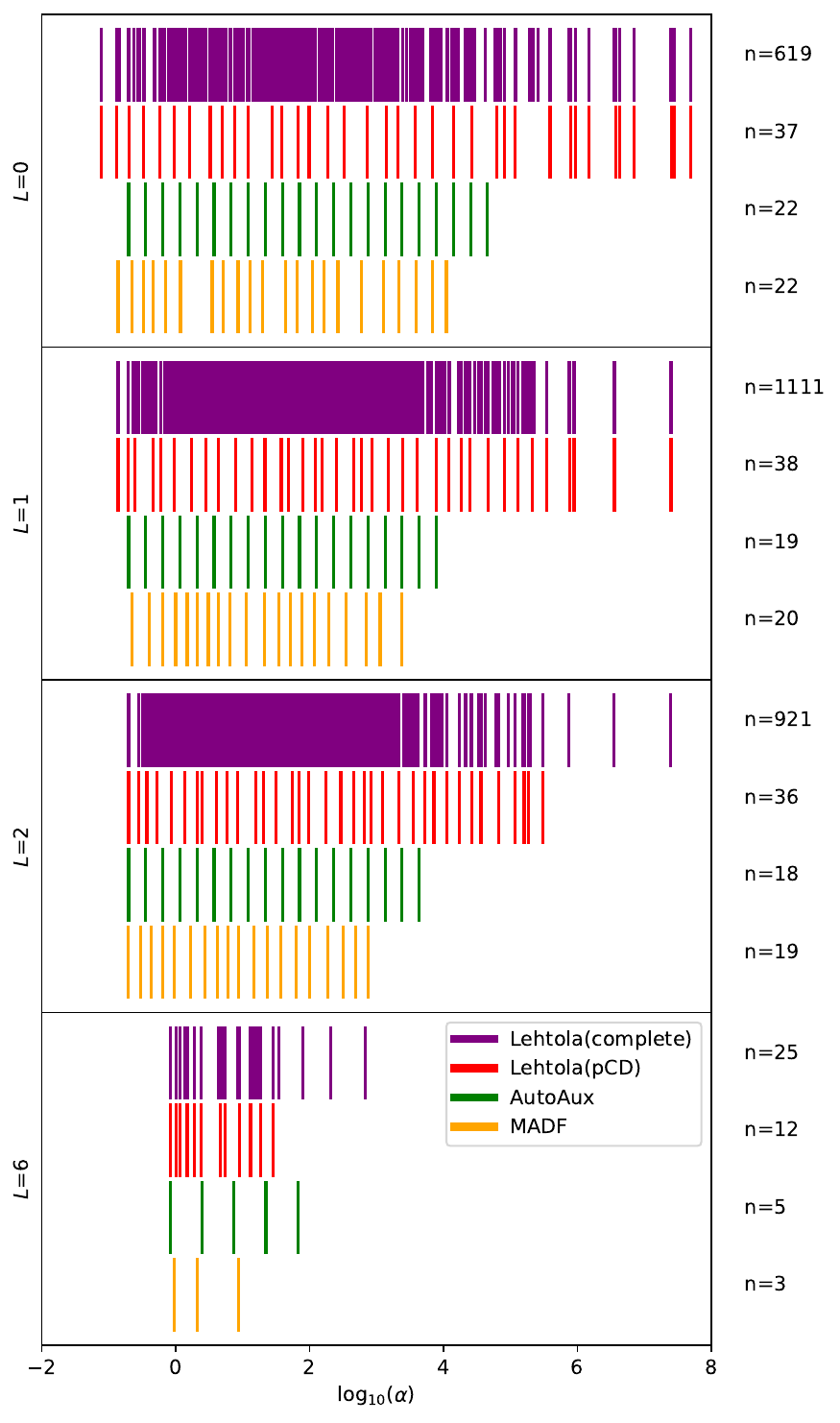}
  \caption{OBS = x2c-QZVPPall-2c}
  \label{fig:rel-exponent-distribution}
\end{subfigure}
\caption{Exponents of DFBS generated (or manually-optimized) for Kr atom with representative nonrelativistic and relativistic OBS using various methods scattered on a logarithmic number line for different angular momenta i.e., $L = \{0, 1, 2, 6 \}$. The number of exponents is shown on the right of each scatter plot. \cref{fig:nonrel-exponent-distribution} shows distribution of DFBS exponents for nonrelativistic OBS def2-QZVPP and \cref{fig:rel-exponent-distribution} for relativistic OBS x2c-QZVPPall-2c. The plot depict Lehtola's complete set (shown in purple), Lehtola's pCD-regularized set (shown in red) obtained using ERKALE\cite{KAS:lehtola:2012:JCC} with pCD threshold $10^{-7}$ as recommended in Ref.\citenum{KAS:lehtola:2021:JCTC},
even-tempered exponents produced by AutoAux (shown in green) and set of exponents produced by the \name algorithm (shown in orange). Distribution of exponents of manually-optimized DFBS def2-QZVPP-RIFIT is also shown (in blue) in \cref{fig:nonrel-exponent-distribution}.}
\label{fig:exponent-distribution}
\end{figure}

Let $\mathcal{C}$ denote the ``complete'' set of all unique solid harmonic primitive AOs generated thereby from all products of primitive OBS AOs on a given atom.
The set $\mathcal{C}$ can be used as DFBS for nearly exact density fitting in an atom (a) for any contraction of the generating set of primitive OBS AOs and (b) for any operator. However, even in a single atom such DFBS is highly linearly dependent and overcomplete, \cite{VRG:lehtola:2019:JCP, KAS:lehtola:2021:JCTC} therefore, it is unusable with finite precision arithmetic.
Moreover, such a basis would be extremely large, defeating the purpose of the DF approximation.
Hence, a compact DFBS is needed to represent the product space of AO functions.
Since the set $\mathcal{C}$ is overcomplete for use as a DFBS for an atom, the unimportant and linearly dependent functions must be trimmed out of this set. 
Lehtola proposed to do such pruning by pCD.\cite{KAS:lehtola:2021:JCTC} 
However, even after pCD the resulting DFBSs are still large, as illustrated in \cref{fig:exponent-distribution}.
While pCD reduces the density of the complete set, it largely leaves its range of exponents intact (at least for lower $L$), and it still keeps many pairs of adjacent exponents with small ratios ($\ll 2$). The problem is especially pronounced for the high angular momentum ($L=6$).

The AutoAux exponent ranges are significantly reduced relative to that of Lehtola's pCD-derived DFBS, and the exponents are well separated owing to even-tempering. However, it is clear that the optimal DFBS exponents are not necessarily even-tempered, as the exponent density of the complete and pCD-processed DFBS of Lehtola and that of the manually-optimized DFBS has significant nonuniformity. Ideally, one would also be able to avoid varying the exponent ratios with $L$ for the even-tempering recipe.

In this work, we regularize the complete candidate exponent pool using a simple adaptive algorithm designed to produce non-even-tempered sets that deviate from the input set as little as possible and with exponent ratios as least as large as the target ratio threshold.
Given a primitive SHG set $\mathcal{C}$ and a target exponent ratio threshold $\zeta$, each subset $\mathcal{C}_L$ containing primitives of angular momenta $L$ is regularized by repeatedly replacing the pair of primitive SHG shells with the smallest exponent ratio $\alpha_1/\alpha_2$ by a single primitive SHG shell with the geometric mean exponent $\alpha =\sqrt{\alpha_{1} \alpha_{2}}$. If multiple exponent pairs have a ``soft'' tie for the smallest exponent ratio, the pair with the smallest orbital exponent wins. Regularization stops if no pairs of primitives have an exponent ratio less than $\zeta$.
Clearly, regularization is most easily implemented if $\mathcal{C}_L$ is first sorted by the exponent in descending order.
The algorithm is described in detail in \cref{alg:regularized_complete_set}.
The recommended value of $\zeta$ will be determined heuristically in \cref{sec:param-optimization}.

\begin{algorithm}
\caption{Generation of regularized complete set of DF SHG shells.}
\begin{algorithmic}[1]
\Function{RegularizedCompleteDFBS}{$\mathcal{P}$, $\zeta$}
\Comment{$\mathcal{P}$: primitive OBS shells; $\zeta\geq1$}
\State $\mathcal{C} \gets \Call{CompleteDFBS}{\mathcal{P}}$ 
\Comment{Complete set of DF SHG shells with exponents generated by \cref{eq:eff-exponent}}
\State $L_{\text{max}} \gets \Call{MaxAM}{\mathcal{c}} $ 
\Comment{Maximum angular momentum of shells in $\mathcal{c}$}
\State $\mathcal{R} \gets \{\}$ 
\Comment{Initialize the regularized complete DFBS}
\For{$L \in [0,L_{\text{max}}]$} 
    \State $\mathcal{c} \gets \Call{SelectByAM}{\mathcal{C}, L}$
    \Comment{Shells in $\mathcal{C}$ with angular momentum $L$}
    \If{\Call{Size}{$\mathcal{c}$}$\geq2$}
        \State $\zeta_\text{min} \gets \Call{MaxRatio}{\mathcal{c}}$
        \Comment{smallest exponent ratio}
        \State $I_\text{min} \gets 0$
        \While{$\zeta_\text{min} < \zeta$}
            \State $\mathcal{c} \gets \Call{SortByExponent}{\mathcal{c} }$
             \Comment{Sort in order of decreasing exponents}
             \State $\mathcal{p} \gets \Call{Neighbors}{\mathcal{c} }$
             \Comment{List of 
             \{\textsc{Exponent}($\mathcal{c}[i]$), \textsc{Exponent}($\mathcal{c}[i+1]$), $i$\} tuples}
            \State $\mathcal{p} \gets \Call{SortByExponentRatio}{\mathcal{p} }$ \Comment{Sort in order of increasing exponent ratios} 
            \State $\zeta_\text{min} \gets \mathcal{p}[0,0]/\mathcal{p}[0,1]$
            \State $\mathcal{p} \gets \Call{MinRatioTies}{\mathcal{p},\zeta_\text{min}}$ \Comment{Select pairs with exponent ratios in $[\zeta_\text{min},\zeta_\text{min}+10^{-8}]$}
            \State $\{ \alpha_I, \alpha_{I+1}, I_\text{min} \} \gets \Call{MinElement}{\mathcal{p}} $
            \Comment{Find the pair with the smallest exponents}
         \State $c \gets \Call{SHGShell}{\sqrt{\alpha_I \alpha_{I+1}},L}$
         \Comment{Pair of shells with closest exponents fused to single shell}
         \State $\mathcal{c} \gets $ \Call{Concat}{$\mathcal{c}\left[0:I_\text{min}-1\right], c, \mathcal{c}\left[I_\text{min}+2:-1\right] $} 
        \EndWhile
    \EndIf
    \State $\mathcal{R} \gets \Call{Concat}{\mathcal{R}, \mathcal{c}} $
\EndFor
\State $\mathcal{R} \gets \Call{CholeskyScreen}{\mathcal{R}, 10^{-8}}$
\Comment{See Ref.\citenum{KAS:lehtola:2021:JCTC}, only drops strongly linearly dependent primitives.}
\State \Return $\mathcal{R}$
\EndFunction
\end{algorithmic}
\label{alg:regularized_complete_set}
\end{algorithm}

\subsection{Estimating energy contribution of DFBS primitives}\label{sec:estimating-2bE-DF-contribution}

In a single atom, DFBS must contain solid harmonics with $L$ up to $2 L_\text{OBS}$ to obtain exact values for all possible matrix elements of the Hamiltonian in OBS, and the above regularization methods do not change this fact. However, it is well-known that the importance of DF AOs with the highest angular momenta is relatively low, at least for energies. Manually-optimized DFBSs typically restrict the $L_\text{DFBS}$ to significantly below $2L_{\text{OBS}}$, since only DFBS with $L$ up to $L_\text{OBS} + L_\text{occ}$ are needed for exact representation of $(ov|ov)$ integrals needed for the MP2 energy. In practice MP2-targeted DFBS are used successfully in the context of higher-order methods, like coupled-cluster. In ad hoc DFBS generators like AutoAux, the angular momenta of DFBS are restricted heuristically, i.e. $L_{\text{max,DF}} = \text{max}(2L_{\text{occ}}, L_{\text{OBS}} + L_{\text{inc}})$, where $L_{\text{inc}}$ is fixed to 1 or 2 depending on $Z$.\cite{KAS:stoychev:2017:JCTC} To avoid the heuristics as much as possible, we attempted to design a first-principles energy-based model for further pruning of DFBS.

Consider the expectation value of a 2-body Fock-space Hamiltonian:
\begin{equation}\label{eq:Rayleigh-Schrödinger-quotient-rdm}
    E =  \sum_{pq} h_{p}^q \gamma_q^p + \frac{1}{2} \sum_{pqrs} h_{pq}^{rs} \gamma_{rs}^{pq},
\end{equation}
where $h_{p}^q$ and $h_{pq}^{rs}$ are the Hamiltonian matrix elements and $\gamma^p_q$ and $\gamma_{rs}^{pq}$ are the elements of the 1- and 2-body reduced density matrices, respectively. Since DF is used to approximate the Coulomb integrals $h_{pq}^{rs} \equiv (pr|qs) \equiv (pr|\hat{J}|qs)$, we will focus on the two-body part of the energy:
\begin{equation}
\label{eq:2-E}
    ^2E = \frac{1}{2} \sum_{pqrs}
    (pr|qs) \gamma_{rs}^{pq}.
\end{equation}
Our objective is to quantitatively estimate the energetic contribution of a given AO in a given DFBS to $^2E$, using a first-principles model for the 2-RDM. 
The most straightforward approach would be to evaluate $^2E$ with the Coulomb integral approximated via robust DF,
\begin{align}
\left( pr|qs\right) = \sum_{XY} (pr|X) ({\bf J}^{-1})_{XY} (Y|qs),
\end{align}
where $(pq|X)\equiv (pq|\hat{J}|X)$ and $({\bf J})_{XY} \equiv (X|\hat{J}|Y)$.
The magnitude of the change in DF-approximated $^2E$ due to the inclusion of the candidate DF AO in DFBS could be used to gauge its importance for accurate computation of $^2E$.
There are two problems with such an approach. 
positive (electrostatic) and negative (exchange, correlation) contributions. Thus, contributions from a given DF AO to the positive and negative parts of $^2E$ could cancel each other spuriously, causing an artificial omission of such AO from DFBS. Replacing the summands in \cref{eq:2-E} by their magnitudes would resolve the sign issue but would, in effect, change the weights of different components of the energy. Second, it is known that the charge density fitting used to compute the electrostatic energy only requires fitting functions with lower angular momenta; thus, it makes sense to tune the fitting basis to the more challenging exchange and correlation contributions to the energy.
 Therefore, we model the energetic importance of DF AOs by the diagonal exchange-like contribution to $^2E$ obtained from \cref{eq:2-E} by substitution $rs \to qp$:
\begin{align}
\label{eq:2-E-dx}
   ^2E^\text{dx} \equiv & \frac{1}{2} \sum_{pq}
    (pq|qp) \gamma_{qp}^{pq} = - \frac{1}{2} \sum_{pq}
    (pq|qp) \gamma_{pq}^{pq}.
\end{align}
Note that $\gamma_{pq}^{pq} \geq 0$, hence each summand in \cref{eq:2-E-dx} is positive (real orbitals are assumed for brevity). 

Another question is how to obtain the 2-RDM. Our original efforts to model the importance of DFBS focused on heuristic models of RDMs (such as the Fermi-Dirac distribution). Thus in practice we constructed an estimator that requires a model for {\em orbital occupancies} $n_p \equiv \gamma_p^p$ only, rather than the full 2-RDM.
To model the 2-RDM in terms of orbital occupancies, we use
the Cauchy-Schwarz inequalities for the 2-RDM,\cite{VRG:helgaker:2000:}
which for positive $\gamma^{pq}_{pq}$ becomes simply:
\begin{equation}
        \gamma^{pq}_{pq} \leq \sqrt{n_p n_q} .
\end{equation}
This leads to the following bound for $^2E^\text{dx}$ in terms of orbital occupancies:
\begin{align}
    \label{eq:2-barE-dx}
   |^2E^\text{dx}| \equiv - ^2E^\text{dx} \leq & \frac{1}{2} \sum_{pq} 
    (pq|qp) \sqrt{n_p n_q} \equiv ~ ^2\bar{E}^\text{dx}.
\end{align}
This expression is independent of whether the single-particle states are spin-free or spin-orbital.
Since {\em robust} DF guarantees that the error in $(pq|qp) = (pq|pq)$ is positive:
\begin{align}
    (pq|pq) - (\widetilde{pq}|\widetilde{pq}) \overset{\text{robust~DF}}{\geq} 0.
\end{align}
then the robust DF approximation to \cref{eq:2-barE-dx},
\begin{align}
\label{eq:2-barE-X-DF}
^2\tilde{\bar{E}}^\text{dx} \equiv & \frac{1}{2} \sum_{XY} \sum_{pq} (pq|X) ({\bf J}^{-1})_{XY} (Y|pq) \sqrt{n_p n_q },
\end{align}
is guaranteed to be smaller than the exact value.
This gives us the ability to estimate the importance of a given DF AO for modeling $^2\bar{E}^\text{dx}$ by monitoring its effect on $^2\tilde{\bar{E}}^\text{dx}$ as well as the ability to control the overall error in $^2\bar{E}^\text{dx}$.

The angular momentum conservation ensures that $X$ and $Y$ in \cref{eq:2-barE-X-DF} must have the same angular momentum. This makes it convenient to split $^2\tilde{\bar{E}}^\text{dx}$ into its angular momentum components:
\begin{align}
\label{eq:2-tildebarE-X-DF-L}
^2\tilde{\bar{E}}^\text{dx}_L \equiv & \frac{1}{2} \sum_{\substack{XY \\ l_X=l_Y=L}} \sum_{pq} (pq|X) ({\bf J}^{-1})_{XY} (Y|pq) \sqrt{n_p n_q },
\end{align}
where $l_X$ is the angular momentum of DF AO $X$.
Practical tests showed that it is important to represent $^2\tilde{\bar{E}}^\text{dx}_L$ for $L \leq 2L_\text{occ}$ more accurately than the higher $L$ channels. Practical tests also showed that $^2\tilde{\bar{E}}^\text{dx}_L$ grows with $Z$, hence it makes sense to prune DFBS according to the value of $^2\tilde{\bar{E}}^\text{dx}_L$ per unit nuclear charge.

The next question is in which order to consider the candidate DF AOs. To decide the order we rewrite the $^2\tilde{\bar{E}}^\text{dx}_L$ in terms of contributions from individual DF AOs:
\begin{align}
\label{eq:2-tildebarepsilon-X-DF-L}
 ^2\tilde{\bar{\epsilon}}^\text{dx}_{X} \equiv & \frac{1}{2} \sum_{\substack{Y \\ l_Y=l_X}} \sum_{pq} (pq|X) ({\bf J}^{-1})_{XY} (Y|pq) \sqrt{n_p n_q } \\
 ^2\tilde{\bar{E}}^\text{dx}_L = & \sum_{\substack{X \\ l_X=L}} {} ^2\bar{\epsilon}^\text{dx}_{X}
\end{align}
Although $^2\bar{\epsilon}^\text{dx}_{X}$ are not strictly positive, they are nearly so; all negative values are very small (on the order of tens of microhartrees).
For each angular momentum channel $L$ vector $|^2\bar{\epsilon}^\text{dx}_{X}|$ is computed once and sorted in descending order; this defines the order in which DF AOs are considered.

2-body energy pruning of the candidate DF AO pool $\mathcal{R}^{L}$ produced by \cref{alg:regularized_complete_set} proceeds as follows.
If $^2\tilde{\bar{E}}^\text{dx}_L$ from (\cref{eq:2-tildebarE-X-DF-L}) is less than $Z \tau$, where $\tau$ is the target threshold, then all DF AOs of this angular momentum are discarded, else the DF AOs are sorted according to the order of importance provided by $|^2\bar{\epsilon}^\text{dx}_{X}|$. 
The candidate DF AOs are added to the $L$-channel of pruned DFBS, $\mathcal{D}^{L}$, until the difference between $^2\tilde{\bar{E}}^\text{dx}_L$ evaluated with $\mathcal{R}^{L}$ and with $\mathcal{D}^{L}$ 
is smaller than $Z \tau$.
For an atom with ground-state configuration including occupied subshells of angular momenta $\leq L_\text{occ}$ DF channels $L\leq 2L_\text{occ}$ are pruned with $\tau=\tau_1$, the rest are pruned with $\tau=\tau_2$.
For atoms with $L_\text{occ} = 0$, namely H, He, Li and Be, to avoid overpruning of the higher $L$ DF channels, we consider $L_\text{occ}$ to be 1.
The values of $\tau_1$ and $\tau_2$will be determined heuristically in \cref{sec:param-optimization}.

\subsection{Correlated model of orbital occupancies}\label{sec:correlated-ensemble-model}

Our initial instinct was to model  orbital occupancies in \cref{eq:2-barE-dx} by Fermi-Dirac distribution (with heuristics for the temperature).  
Unfortunately such an approach did not prove fruitful. Using atomic mean-field ensemble orbitals often used as initial guess for atomic density matrix would not suffice, since this will fail to generate nonzero populations for orbitals important in dynamical electron correlation.
Therefore we developed a simple model for correlated 1-RDM of a neutral atom in an ensemble state using spin-opposite scaled (SOS) first-order M\o{}ller-Plesset perturbation of a model mean-field ensemble density.
The mean-field density in OBS is obtained by one-shot diagonalization of OBS Fock matrix $\mathbf{F}$ computed from the atomic ensemble density generated by populating subshells of a minimal basis using the atom's reference ground-state electron configuration. Such definition for the OBS Fock matrix is utilized in the \code{Libint} library and the \code{MPQC} program under the name ``superposition of atomic densities'' (SOAD) for generating guess Fock matrices for SCF. SOAD can be viewed as a simplified version of the popular SAD method\cite{VRG:vanlenthe:2006:JCC}, with differences described for completeness in \cref{app:soad}.

Diagonalization of SOAD ensemble Fock matrix expressed in OBS AO (see \cref{app:soad}),
\begin{equation}\label{eq:fc=sce}
    \mathbf{F}_{\mathrm{SOAD}}\mathbf{C} = \mathbf{SC}\boldsymbol{\varepsilon},
\end{equation}
produces energies and OBS AO coefficients of non-self-consistent ensemble orbitals. In this work, we have used the spin-free exact two-component 1-body (core) Hamiltonian (sf-1eX2C)\cite{KAS:filatov:2003:JCP, VRG:kutzelnigg:2005:JCP, VRG:peng:2013:JCP} to construct the $\mathbf{F}_{\mathrm{SOAD}}$ matrix to account for scalar relativistic effects. In general the model orbital energies of the atomic subshells obtained from \cref{eq:fc=sce} do not match the canonical order of Aufbau principle (Madelung rule) or even the order produced by self-consistent field in the given OBS. Thus the uncorrelated occupation numbers for the OBS ensemble orbitals ${\bf n}^{(0)}$ are obtained by mapping the spherically-averaged MBS ensemble occupancies ${\bf n}_\mathrm{SOAD}$ (\cref{app:soad}) via:
\begin{align}
\label{eq:n0}
{\bf n}^{(0)} = \mathrm{id}\left(\mathbf{C}^\dagger \mathbf{S}_\mathrm{OBS,MBS}\right) {\bf n}_\mathrm{SOAD},
\end{align}
where $\mathbf{S}_\mathrm{OBS,MBS}$ is the overlap of OBS and MBS AOs, and $\mathrm{id}$ maps the real-valued matrix argument to the nearest identity matrix.
Columns in ${\bf C}$ corresponding to zero and nonzero ensemble occupancies are deemed unoccupied ($a,b$) and occupied ($i,j$), respectively. These orbitals and the corresponding orbital energies are used to compute opposite-spin ensemble M\o{}ller-Plesset first-order amplitudes:
\begin{equation}\label{eq:sos_t2}
    t_{ab}^{ij} = -\frac{\sqrt{n_i^{(0)} n_j^{(0)}}}{2} \frac{(ia|jb)}{\varepsilon_{a} + \varepsilon_b -\varepsilon_i - \varepsilon_j} (1 - e^{-\kappa (\varepsilon_{a} + \varepsilon_b -\varepsilon_i - \varepsilon_j)  })^2 .
\end{equation}
The last factor on the right-hand side is the $\kappa$-regularizer used by Lee and Head-Gordon\cite{KAS:lee:2018:JCTC} in the context of orbital-optimized MP2 method. Whereas in Ref. \citenum{KAS:lee:2018:JCTC} heuristic tuning suggested $\kappa=1.45 \, E_\mathrm{h}^{-1}$ as best at avoiding the singularities and accounting for the inaccuracy of the MP1 amplitude model. Here we set $\kappa=3 \, E_\mathrm{h}^{-1}$ as the purpose of the regularizer here is to only avoid potential singularities due to possible degeneracies between occupied and unoccupied subshells; the inadequacy of the MP1 model for describing correlations of electrons in the partially occupied subshells is already partially addressed by avoiding internal excitations within such subshells.
The corresponding second-order correlation contributions ${\bf n}^{(2)}$ to orbital occupancies are obtained straightforwardly:
\begin{align}\label{eq:sos-1rdm}
      n_i^{(2)} = & \sum_{jab} -2t^{ij}_{ab} t_{ij}^{ab},\\
      n_a^{(2)} = & \sum_{ijb} 2t^{ij}_{ab} t_{ij}^{ab}.
\end{align}
The total correlated orbital occupancy vector ${\bf n}$ is a sum of the mean-field and correlated contributions:
\begin{align}\label{eq:total-1rdm}
    {\bf n} = & {\bf n}^{(0)} + {\bf n}^{(2)}.
\end{align}

\subsection{The \name Algorithm}\label{sec:model}

For a given (input) OBS consisting of contracted SHG AOs on an atom with atomic number $Z$ \name (1) generates a ``complete'' set of primitive DF SHG AOs necessary to represent all products of uncontracted OBS AOs using the method of Lehtola\cite{KAS:lehtola:2021:JCTC}, (2) regularizes it by fusing pairs of shells with overly close exponents as described in \cref{sec:products-of-gaussians}, and (3) prunes out DF AOs not important for description of correlated 2-body energy using the method described in \cref{sec:estimating-2bE-DF-contribution}. The complete description of the algorithm is given in \cref{alg:madf}. Its 3 model parameters (with 4 unique values) are summarized in \cref{tab:madf-params}. 

\begin{table}[h]
    \begin{threeparttable}
    \centering
    \begin{tabular}{c|c|c}
    \hline\hline
    Parameter & Description & Values\\ \hline
      $\zeta$   &  Exponent ratio threshold for \cref{alg:regularized_complete_set} & 1.4\\
      $\tau_1$   &   2-body energy threshold for $L\leq 2l^{occ}$ & $10^{-6}$ (nr),  $10^{-7}$ (rel) \tnote{a}\\
      $\tau_2$   &  2-body energy screening threshold for $L > 2l^{occ}$ & $10^{-5}$ \\
         \hline\hline
    \end{tabular}
    \caption{Model parameters of the \name DFBS generator.}
    \begin{tablenotes}
    \item[a] {Optimal values of $\tau_1$ are different for nonrelativistic and relativistic OBSs, see \cref{sec:result-nonrel-param,sec:result-rel-param}}
    \end{tablenotes}
    \label{tab:madf-params}
    \end{threeparttable}
\end{table}

\begin{algorithm}
\caption{The \name algorithm}
\begin{algorithmic}[1]

\Function{\name}{$Z, \mathcal{O}, \mathcal{M}, \tau_1 , \tau_2  , \zeta $} 
\Comment{$Z$: atomic number; $\mathcal{O}$: (contracted) OBS; $\mathcal{M}$: MBS; $\tau_{\{1,2\}}\geq0$; $\zeta\geq1$}
\State $\mathcal{P} \gets \Call{Uncontract}{\mathcal{O}}$
\State $ \mathcal{C} \gets \Call{RegularizedCompleteDFBS}{\mathcal{P}, \zeta}$ \Comment{\cref{alg:regularized_complete_set}}
\State $\mathbf{F}_{\mathrm{SOAD}} \gets \Call{SoadFock}{Z, \mathcal{O}, \mathcal{M}}$
\Comment{\cref{app:soad}}
\State $\mathbf{C} \gets \mathbf{F}_{\mathrm{SOAD}}\mathbf{C} = \mathbf{SC\epsilon}$ \Comment{\cref{eq:fc=sce}}
\State ${\bf n} \gets \Call{CorrelatedOccupancies}{\mathbf{C}, \mathcal{M}}$ 
\Comment{\cref{eq:n0,eq:sos_t2,eq:sos-1rdm,eq:total-1rdm}}
\State $L_{\text{occ}} \gets \Call{Max}{\textsc{MaxAM}(\mathcal{M}), 1}$
\Comment{maximum angular momentum of occupied subshells.}
\State $L_{\text{max}} \gets \Call{MaxAM}{\mathcal{R}}$ 
\Comment{maximum angular momentum of candidate pool of shells}
\State $\mathcal{D} \gets \{ \} $ 
\Comment{Initialize the target list of DF shells}
\For{$L \in [0,L_{\text{max}}]$}
    \State $\mathcal{d} \gets \{ \}$
    \If{$L\leq 2 L_{\text{occ}}$}
        \State $\tau \gets \tau_1$
    \Else
        \State $\tau \gets \tau_2$
    \EndIf

    \State $\mathcal{r} \gets \Call{SelectByAM}{\mathcal{R}, L}$
    \Comment{Shells in $\mathcal{R}$ with angular momentum $L$}
    \State $\{E_\mathcal{r}, \{\epsilon_\mathcal{r}\}\}  \gets \Call{Eval2EdxL}{\mathcal{r}}$
    \Comment{Evaluate  \cref{eq:2-tildebarE-X-DF-L,eq:2-tildebarepsilon-X-DF-L} using $\mathcal{r}$ as DFBS}
    \If{$E_\mathcal{r} \geq Z \tau$}
        \State $\mathcal{r} \gets \Call{SortBy}{\mathcal{r},\{|\epsilon_\mathcal{r}|\}}$ 
        \Comment{Sort $\mathcal{r}$ in order of decreasing $|{}^2\tilde{\bar{\epsilon}}^\text{dx}_{X}|$}
        \State $ E_\mathcal{d} \gets 0, i \gets 0$
        \While{$E_\mathcal{r} - E_\mathcal{d} \geq Z \tau$ and $i<\Call{Size}{r}$} 
            \State $\mathcal{d} \gets \Call{Concat}{\mathcal{d}, \mathcal{r}[i]} $
            \State $\{E_\mathcal{d}, \_ \}  \gets \Call{Eval2EdxL}{\mathcal{d}}$
    \Comment{Evaluate  \cref{eq:2-tildebarE-X-DF-L} using $\mathcal{d}$ as DFBS}
            \State $i \gets i+1$
        \EndWhile
        \State $\mathcal{D} \gets \Call{Concat}{\mathcal{D}, \mathcal{d}} $ 
    \EndIf
\EndFor
\State \Return $\mathcal{D}$
\EndFunction
\end{algorithmic}
\label{alg:madf}
\end{algorithm}

\section{Technical details}\label{sec:technical}
The \name algorithm is implemented in a developmental version of \code{MPQC}\cite{VRG:peng:2020:JCP};
generation of base DFBS by Lehtola's algorithm\cite{KAS:lehtola:2021:JCTC} was implemented in \code{Libint}.\cite{Libint2,libint-2.9.0}
For optimization of \name parameters and its benchmark testing, we employed the nonrelativistic and relativistic Hartree-Fock (HF) and second-order M\o{}ller-Plesset (MP2) methods. 
 DF errors were computed relative to the HF and MP2 energies obtained with exact (four-center) two-electron matrix elements throughout.
The exact two-component decoupled one-body Hamiltonian\cite{KAS:filatov:2003:JCP, VRG:kutzelnigg:2005:JCP, VRG:peng:2013:JCP} (1eX2C, or simply X2C) with spin-orbit coupling was used for all relativistic computations without empirical scaling corrections.
The core orbitals were frozen in nonrelativistic MP2 computations with cc-pVXZ and def2- OBSs.

The \name model parameters for use with nonrelativistic OBS were determined on training set TS1 composed of 38 closed-shell systems with light elements: AlF$_3$, Ar, BF$_3$, BH$_3$, C$_2$H$_2$, C$_2$H$_4$, C$_2$H$_6$, CH$_3$OH, CH$_4$, Cl$_2$, CO$_2$, COH$_2$, CS$_2$, F$_2$, H$_2$O, HCl, HF, HNO$_2$, N$_2$, cis-N$_2$H$_2$, trans-N$_2$H$_2$, Ne, NH$_3$, NOCl, O$_2$, O$_3$, OF$_2$, P$_2$, PF$_3$, PH$_3$, S$_2$, SF$_2$, SiH$_4$, SiO$_2$, and  SO$_2$. The performance of the recommended model parameters was assessed on the standard G2 set of molecules\cite{KAS:curtiss:1997:JCP}, containing 118 closed- and 30 open-shell systems.
For training and assessment computations cc-pVXZ\cite{VRG:dunning:1989:JCP}, cc-pwCVXZ\cite{KAS:peterson:2002:JCP}(X = D, T, Q) and def2-SVP, def2-XZVP (X = T, Q)\cite{KAS:weigend:2005:PCCP} were used.
Since there are no cc-pwCVXZ basis sets for Li, Be, and Na atoms, systems containing these atoms were skipped from the assessment of these OBSs.

The \name model parameters for use with relativistic OBS were determined on training set TS2 composed of 24 closed-shell systems with heavy elements: AgH, AsH$_3$, AuH, BiH$_3$, CuH, GaH$_3$, GeH$_4$, HAt, HBr, Hi, InH$_3$, Kr, PbH$_4$, PoH$_2$, PtC, RuC, SbH$_3$, ScH, SeH$_2$, SnH$_4$, TeH$_2$, TlH$_3$, Xe and Rn. 
The performance of the recommended model parameters was assessed on 2 sets: (a) the Ln54 benchmark set  composed of 54 rare-earth compounds\cite{KAS:aebersold:2017:JCTC}, and (b) the set of 60 diatomics containing d-block (transition metal) atoms from Ref. \citenum{KAS:aoto:2017:JCTC} (we will refer to this set as Tm60).
Due to the challenging SCF orbital optimization landscape and to make the assessment as robust and automated, for each OBS we did assessment on the Ln54 and Tm60 systems for which the exact (four-center) X2C-HF SCF converged in 30 or fewer SCF iterations.
The training and assessment was performed with the following OBSs designed for spin-orbit relativity: X2C-SVPall-2c\cite{VRG:pollak:2017:JCTC}, X2C-XZVPPall-2c\cite{KAS:franzke:2020:JCTC} (X = T, Q) and  all-electron Dyall basis sets dyall-aeYz\cite{VRG:gomes:2010:TCA,KAS:dyall:2012:TCA,KAS:dyall:2023:} (Y = 2, 3, 4). 

The performance of \name was assessed also against the AutoAux DFBS generator\cite{KAS:stoychev:2017:JCTC}
as implemented in BSE's command-line interface\cite{VRG:pritchard:2019:JCIM} .
For cc-pVXZ, cc-pwCVXZ, and def2- family of OBSs we also compared against the corresponding manually-optimized DFBS cc-pVXZ-RIFIT, cc-pwCVXZ-RIFIT (X = D, T, Q)  and def2-SVP-RIFIT and def2-XZVP-RIFIT (X = T, Q) for the G2 set for comparison.
acCD calculations were performed using \code{OpenMolcas} version 24.10\cite{KAS:limanni:2023:JCTC}. Since this version of \code{OpenMolcas} does not have an implementation of open-shell MP2, for comparison of \name and acCD only the closed-shell subset of G2 set was used. We also show acCD vs \name comparison only for the cc-pVXZ family of basis sets. 
All geometries, except for the G2 set, were optimized with the PBE0\cite{KAS:adamo:1999:JCP} and def2-TZVP basis set using the \code{Psi4}'s\cite{VRG:smith:2020:JCP} geometry optimization module.
For elements with atomic number greater than 36, the effective core potentials (ECPs)\cite{KAS:cao:2011:WCMS} available for def2-TZVP were used for geometry optimization. 

All matrix pseudo-inverses and pseudo-inverse square roots were computed with L\"o{}wdin's orthogonalization procedure\cite{KAS:lowdin:1950:JCP}.

\section{Results}\label{sec:results}

\subsection{Training \name model parameters}\label{sec:param-optimization}

The \name model parameters were tuned to ensure that that the DF errors for the respective training sets did not exceed target accuracies of $\pm 20 \mu E_h$ and $\pm 10 \mu E_h$ per electron for HF and MP2 energies, respectively. This target range for errors was determined by analyzing the DF errors of manually-optimized DFBSs; the majority have larger errors than these targets (see \cref{sec:result-g2}).

\subsubsection{Training \texorpdfstring{$\tau_{1,2}$ and $\zeta$}{tau1,2 and zeta} for nonrelativistic OBS}\label{sec:result-nonrel-param}

\begin{figure}[htp]
\begin{subfigure}{\textwidth}
    \centering
    \includegraphics[width=0.8\textwidth]{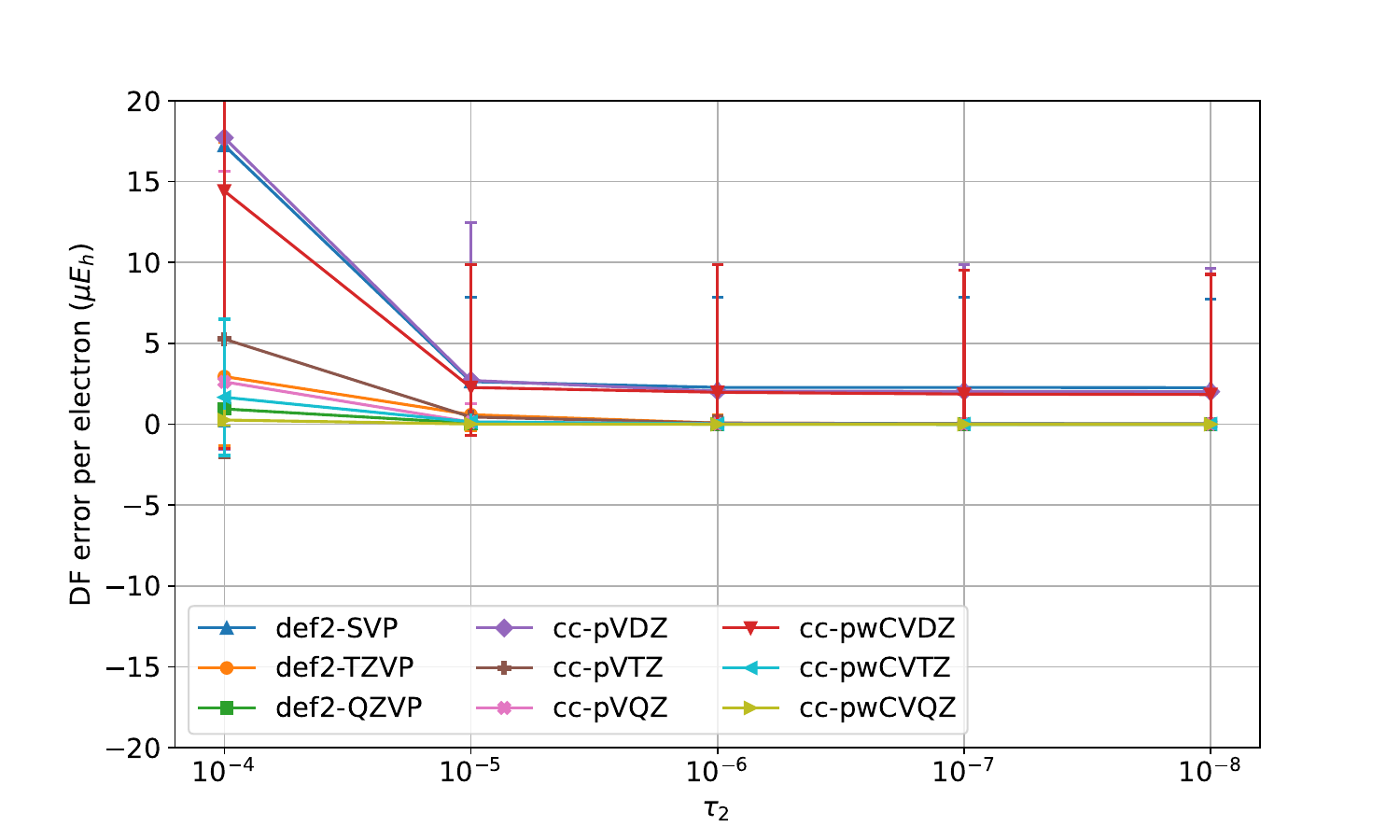}
    \caption{}
    \label{fig:tau2-hf}
\end{subfigure}
\bigskip
\begin{subfigure}{\textwidth}
    \centering
    \includegraphics[width=0.8\textwidth]{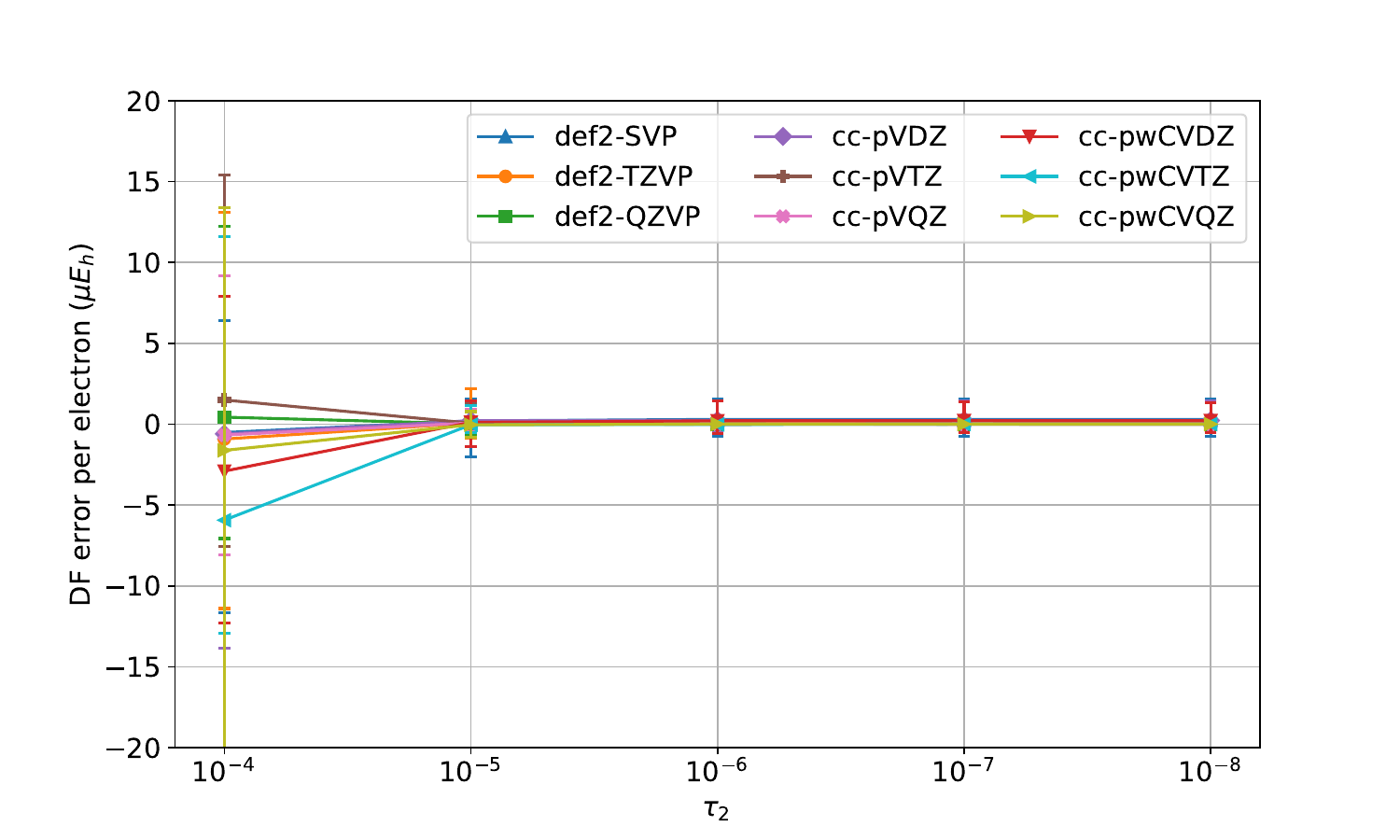}
    \caption{}
    \label{fig:tau2-mp2}
\end{subfigure}
\caption{Variation of DF errors of nonrelativistic (a) HF and (b) MP2  energies of the TS1 training set vs the $\tau_2$ model parameter of the \name generator.}
\end{figure}

\begin{figure}[htp]
\begin{subfigure}{\textwidth}
    \centering
    \includegraphics[width=0.8\textwidth]{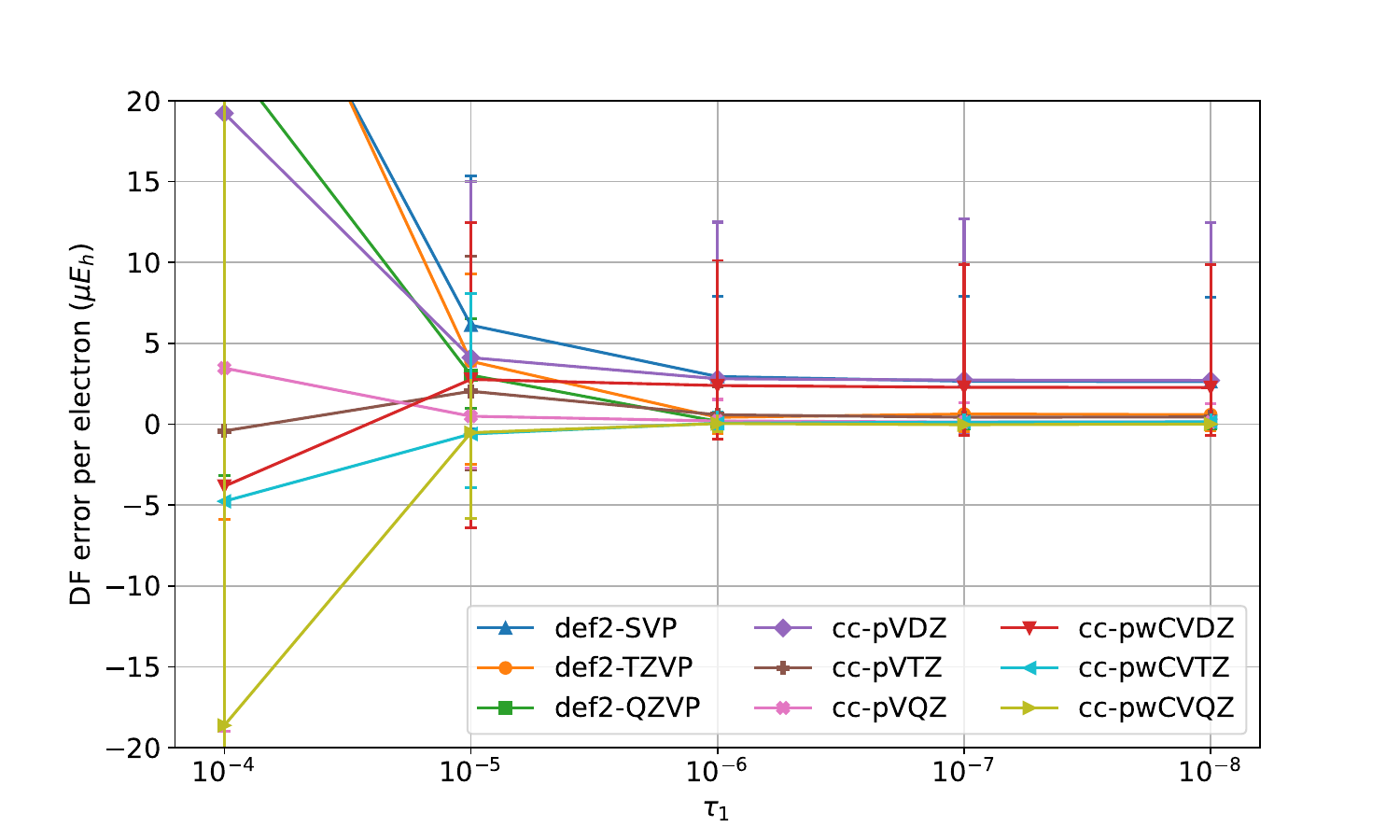}
    \caption{}
    \label{fig:tau1-hf}
\end{subfigure}
\bigskip
\begin{subfigure}{\textwidth}
    \centering
    \includegraphics[width=0.8\textwidth]{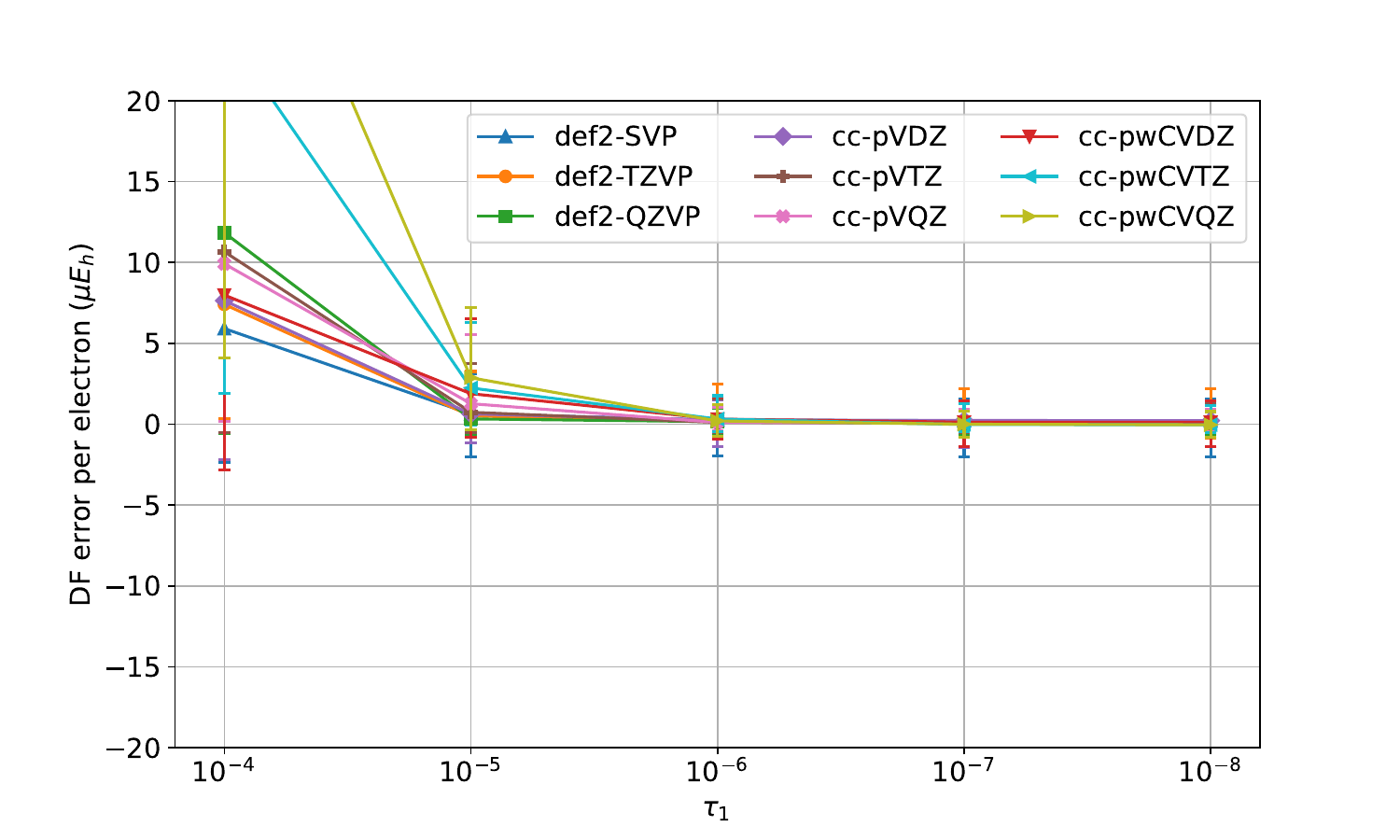}
    \caption{}
    \label{fig:tau1-mp2}
\end{subfigure}
\caption{Variation of DF errors of nonrelativistic (a) HF and (b) MP2  energies of the TS1 training set vs the $\tau_1$ model parameter of the \name generator.}
\end{figure}

\begin{figure}[htp]
\begin{subfigure}{\textwidth}
    \centering
    \includegraphics[width=0.8\textwidth]{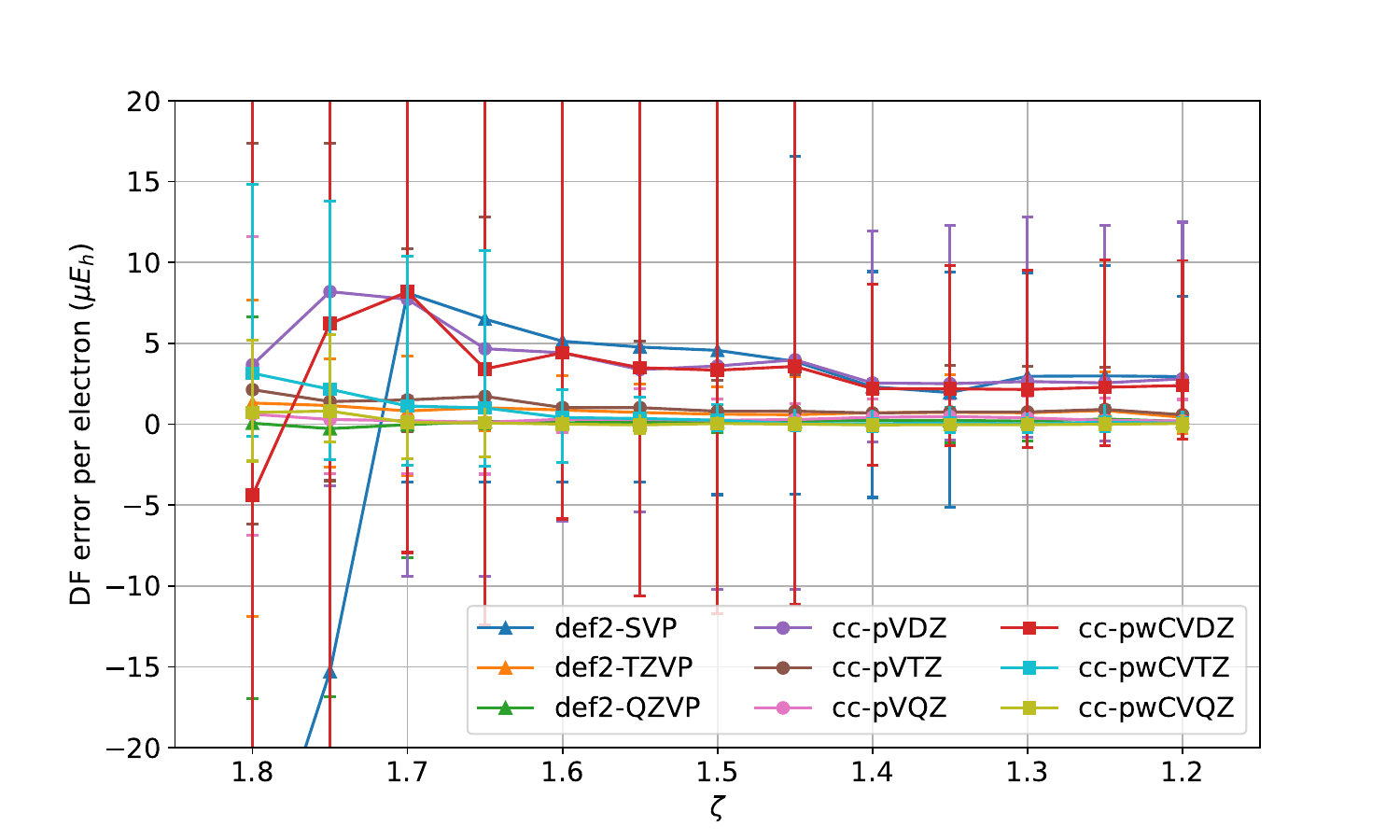}
    \caption{}
    \label{fig:zeta-hf}
\end{subfigure}
\bigskip
\begin{subfigure}{\textwidth}
    \centering
    \includegraphics[width=0.8\textwidth]{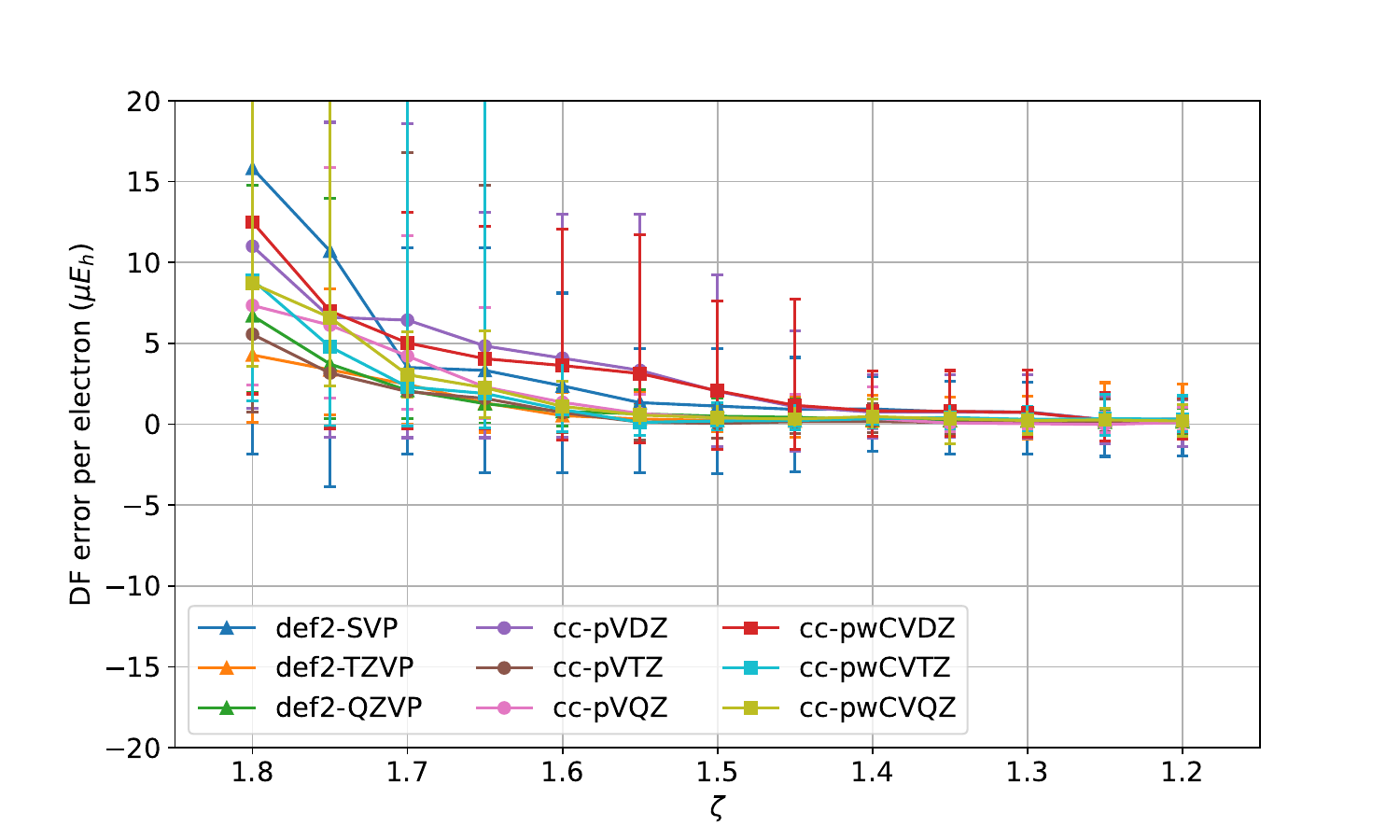}
    \caption{}
    \label{fig:zeta-mp2}
\end{subfigure}
\caption{Variation of DF errors of nonrelativistic (a) HF and (b) MP2  energies of the TS1 training set vs the $\zeta$ model parameter of the \name generator.}
\end{figure}

The variation of DF errors with $\tau_2$ was studied next, by keeping the rest of the parameters fixed ($\tau_1=10^{-8}, \zeta=1.2$) sufficiently closely to their asymptotic limits ($\tau\to0$, $\zeta\to1$) without causing ill-conditioning and excessive costs.
As \cref{fig:tau2-hf,fig:tau2-mp2} indicate, the DF errors of HF and MP2 are largely converged at $\tau_2=10^{-5}$.
Next, $\tau_2$ was fixed at $10^{-5}$ and $\tau_1$ was varied
(\cref{fig:tau1-hf,fig:tau1-mp2}). The DF errors are sufficiently converged with $\tau_1=10^{-6}$.
Finally, with $\tau_1 = 10^{-6}$ and $\tau_2=10^{-5}$ variation of DF errors with $\zeta$ was studied (\cref{fig:zeta-hf,fig:zeta-mp2}). For $\zeta > 1.4$ the max DF errors increase due to the insufficient spanning of the OBS AO product space by the DFBS AOs, whereas for $\zeta<1.2$ (not shown) the onset of ill-conditioning in DFBS makes the \name model not numerically stable. $\zeta=1.4$ ensures sufficient convergence of the DF errors without undue numerical problems.

Note that $\zeta=1.4$ is significantly smaller than the exponent ratio 1.8 used by the AutoAux generator ( 1.8)\cite{KAS:stoychev:2017:JCTC} and ratio 2 used by \code{PySCF}'s generator\cite{pu2025enhancingpyscfbasedquantumchemistry} for even-tempered DFBS construction. Our findings confirm that the relatively small exponent ratios are indeed necessary for accurate density fitting. For example, exponent ratios in the $[1.3,1.5]$ range are quite common in the $L \leq 2$ channels of the manually-optimized def2-QZVPP-RIFIT DFBS (see \cref{fig:exponent-distribution}). Clearly even-tempered spanning of the AO product space is suboptimal for practical basis sets.

\subsubsection{Retraining \name parameters for relativistic OBS.}\label{sec:result-rel-param}
\begin{figure}[htp]
\begin{subfigure}{\textwidth}
    \centering
    \includegraphics[width=0.8\textwidth]{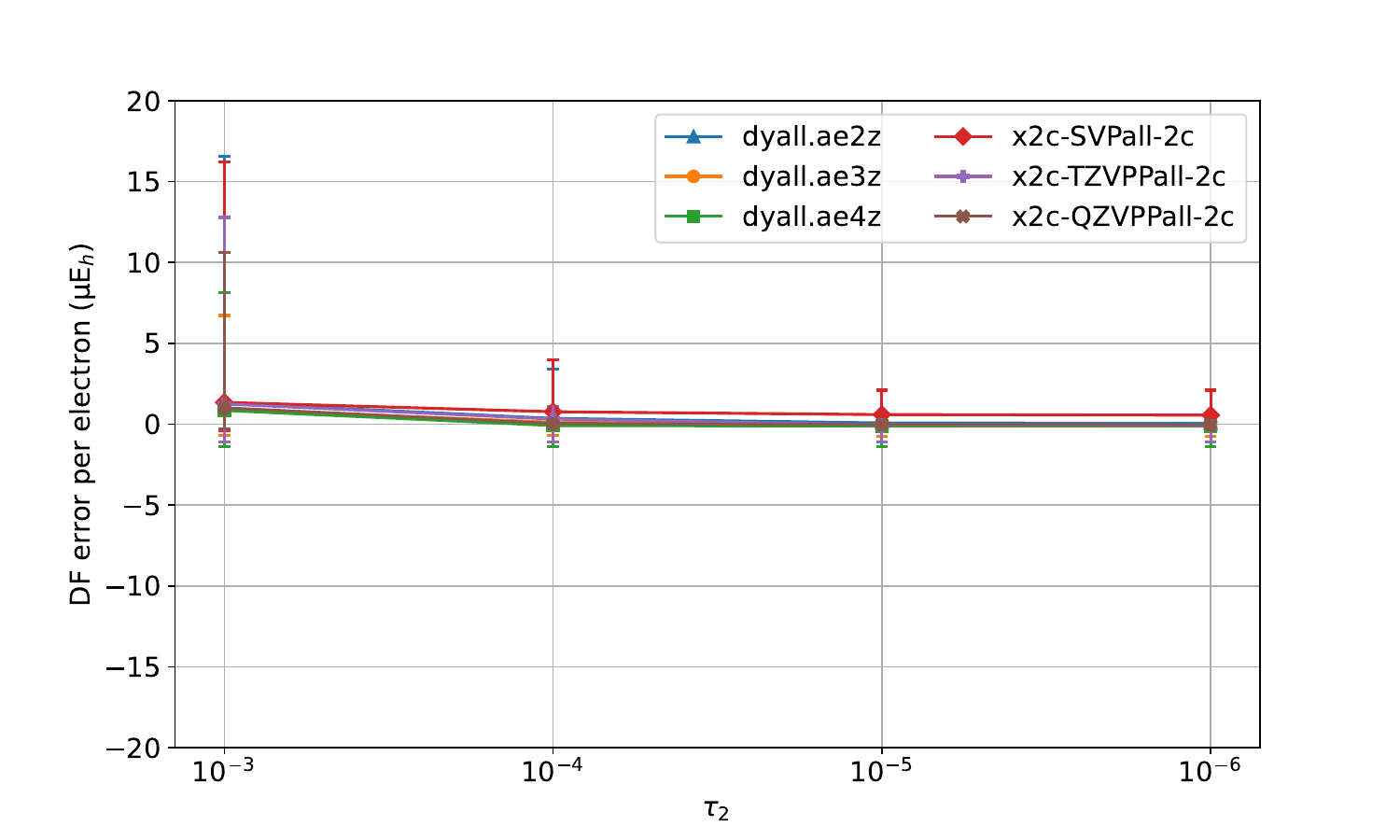}
    \caption{}
    \label{fig:rel-tau2-hf}
\end{subfigure}
\bigskip
\begin{subfigure}{\textwidth}
    \centering
    \includegraphics[width=0.8\textwidth]{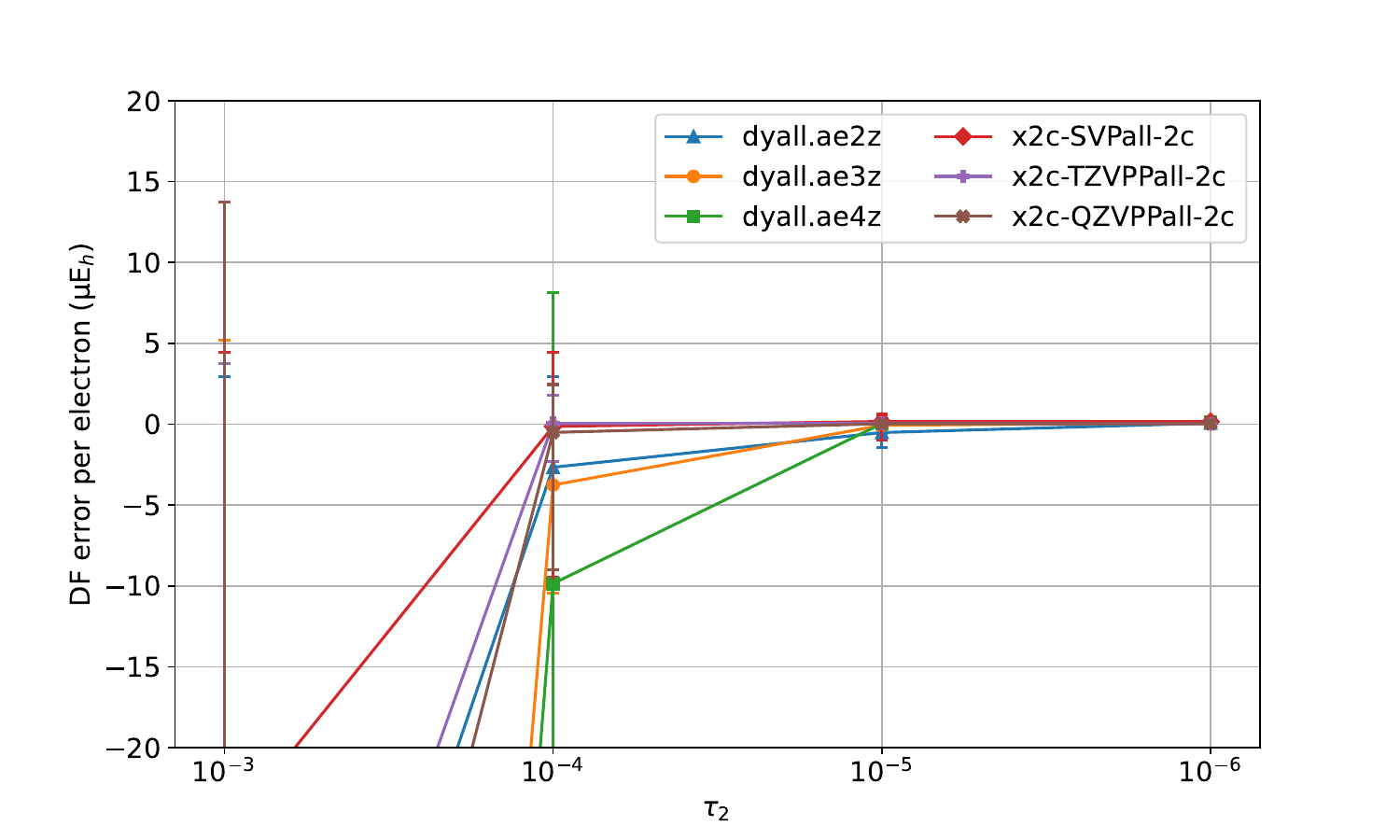}
    \caption{}
    \label{fig:rel-tau2-mp2}
\end{subfigure}
\caption{Variation of DF errors of (relativistic) (a) X2C-HF and (b) X2C-MP2 energies of the TS2 training set vs the $\tau_2$ model parameter of the \name generator.}
\end{figure}

\begin{figure}[htp]
\begin{subfigure}{\textwidth}
    \centering
    \includegraphics[width=0.8\textwidth]{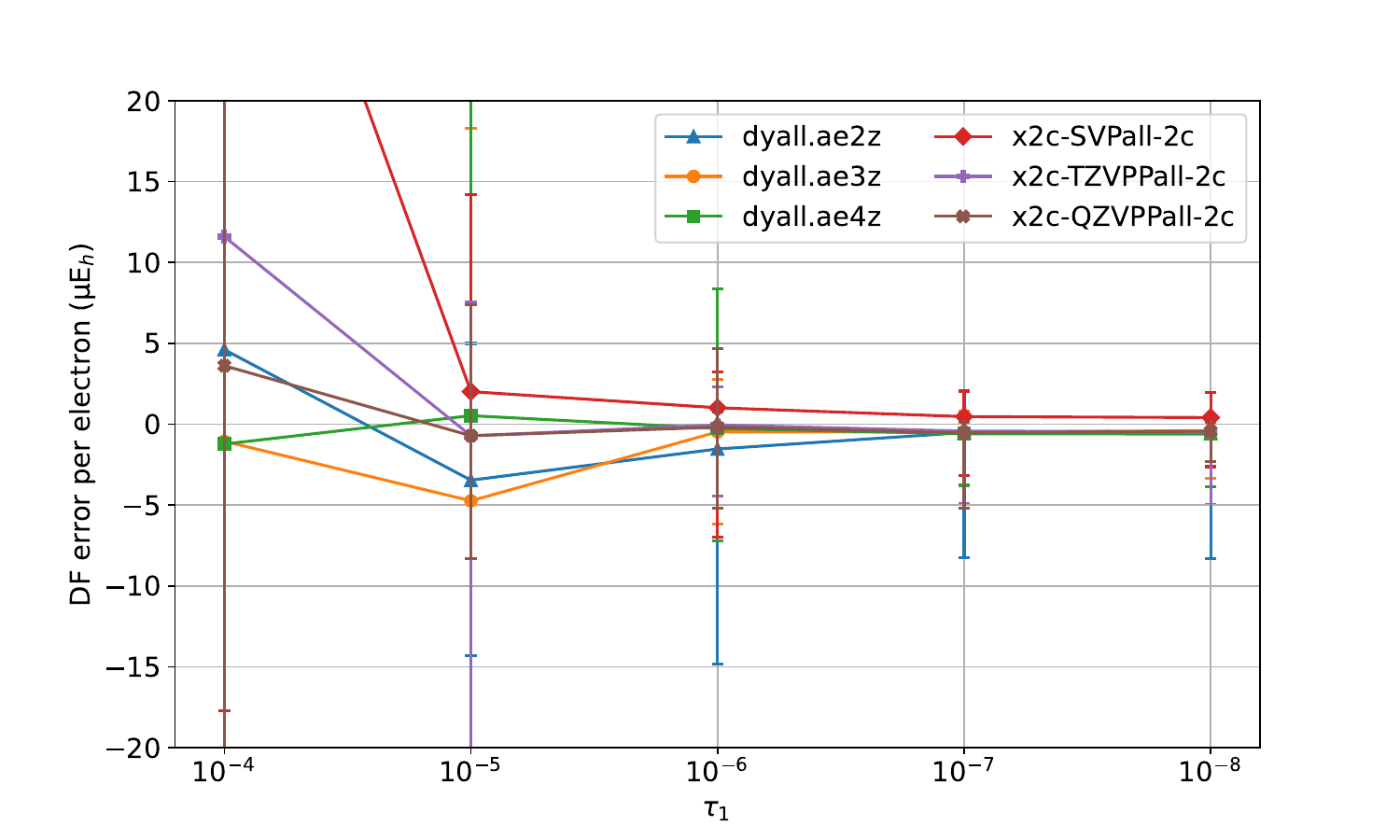}
    \caption{}
    \label{fig:rel-tau1-hf}
\end{subfigure}
\bigskip
\begin{subfigure}{\textwidth}
    \centering
    \includegraphics[width=0.8\textwidth]{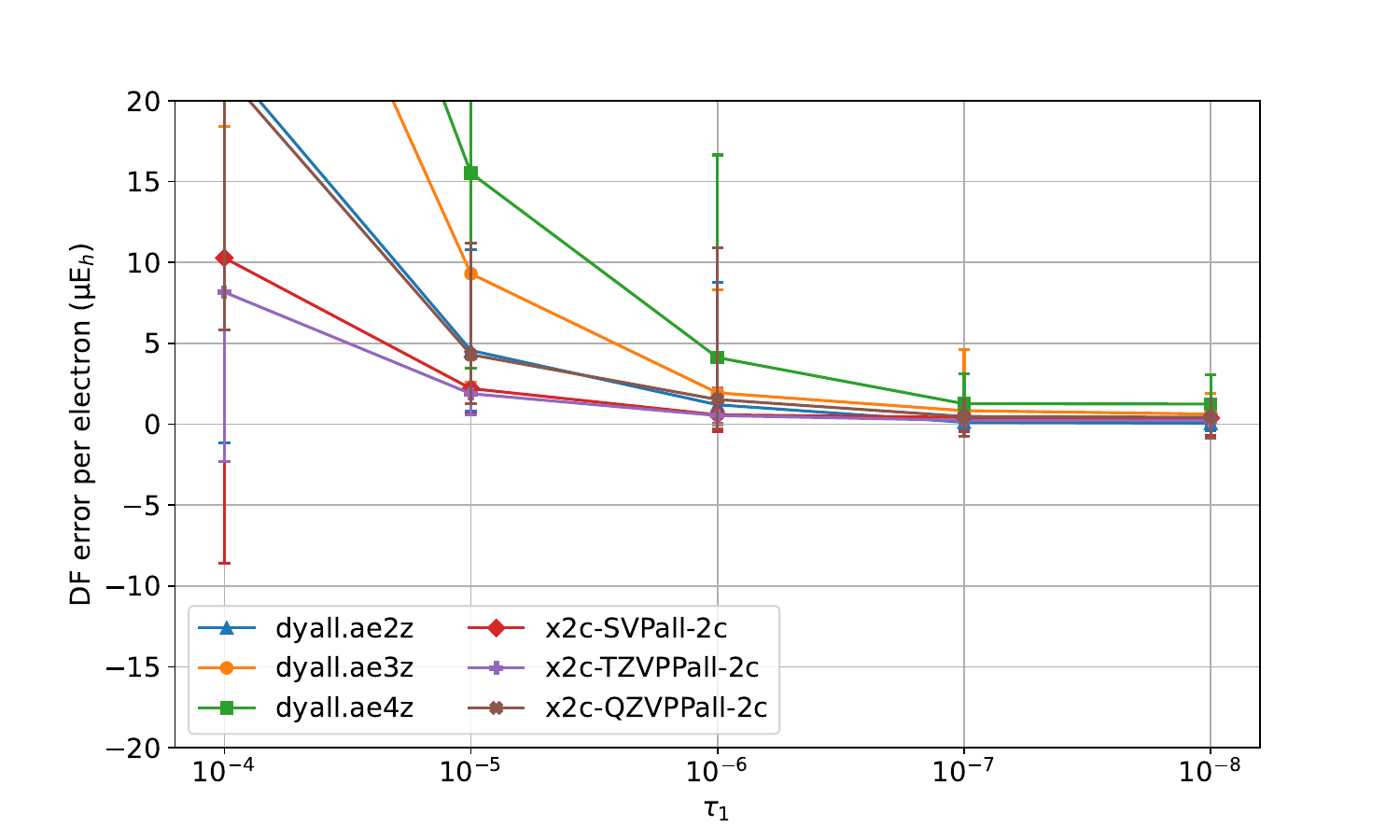}
    \caption{}
    \label{fig:rel-tau1-mp2}
\end{subfigure}
\caption{Variation of DF errors of (relativistic) (a) X2C-HF and (b) X2C-MP2 energies of the TS2 training set vs the $\tau_1$ model parameter of the \name generator.}
\end{figure}

\begin{figure}[htp]
\begin{subfigure}{\textwidth}
    \centering
    \includegraphics[width=0.8\textwidth]{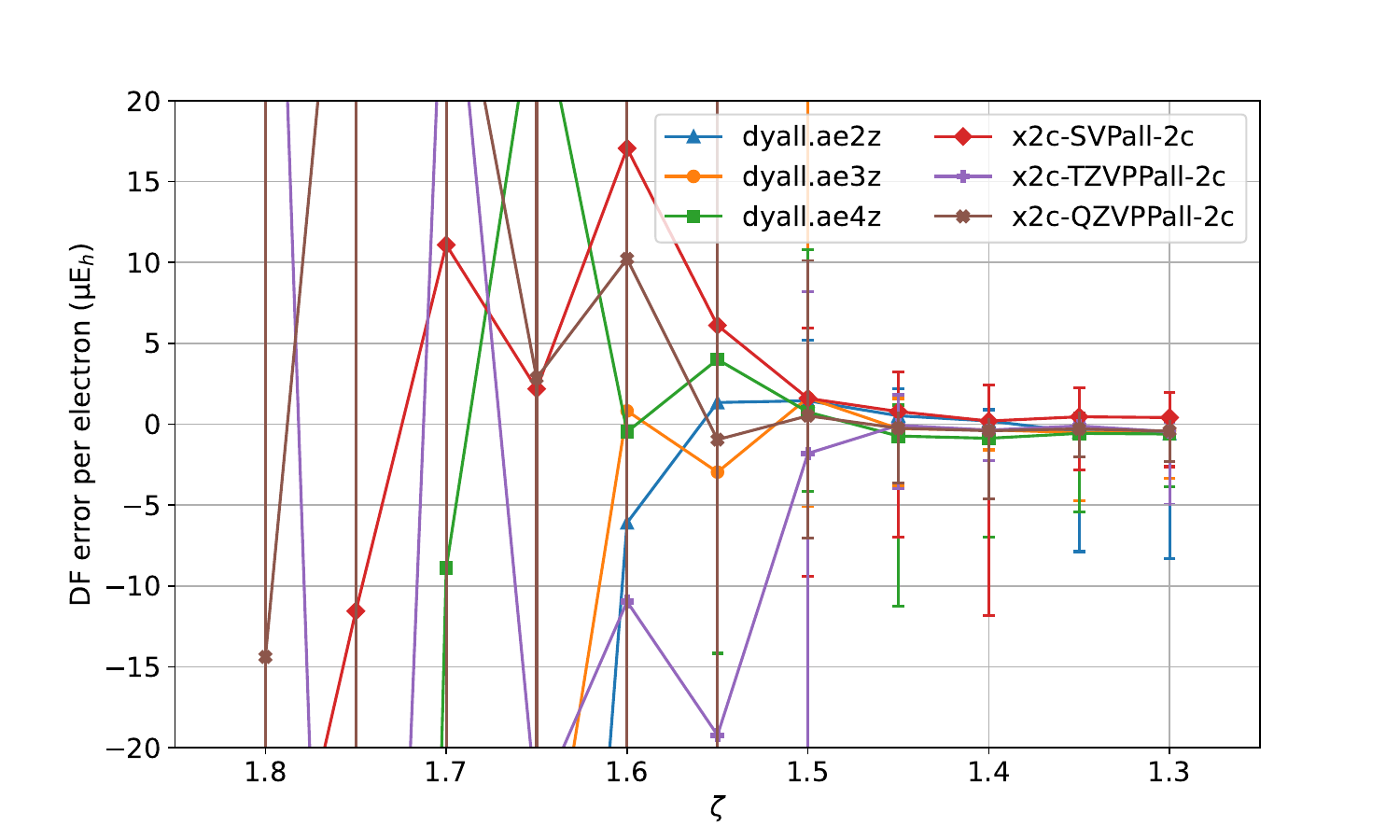}
    \caption{}
    \label{fig:rel-zetas-hf}
\end{subfigure}
\bigskip
\begin{subfigure}{\textwidth}
    \centering
    \includegraphics[width=0.8\textwidth]{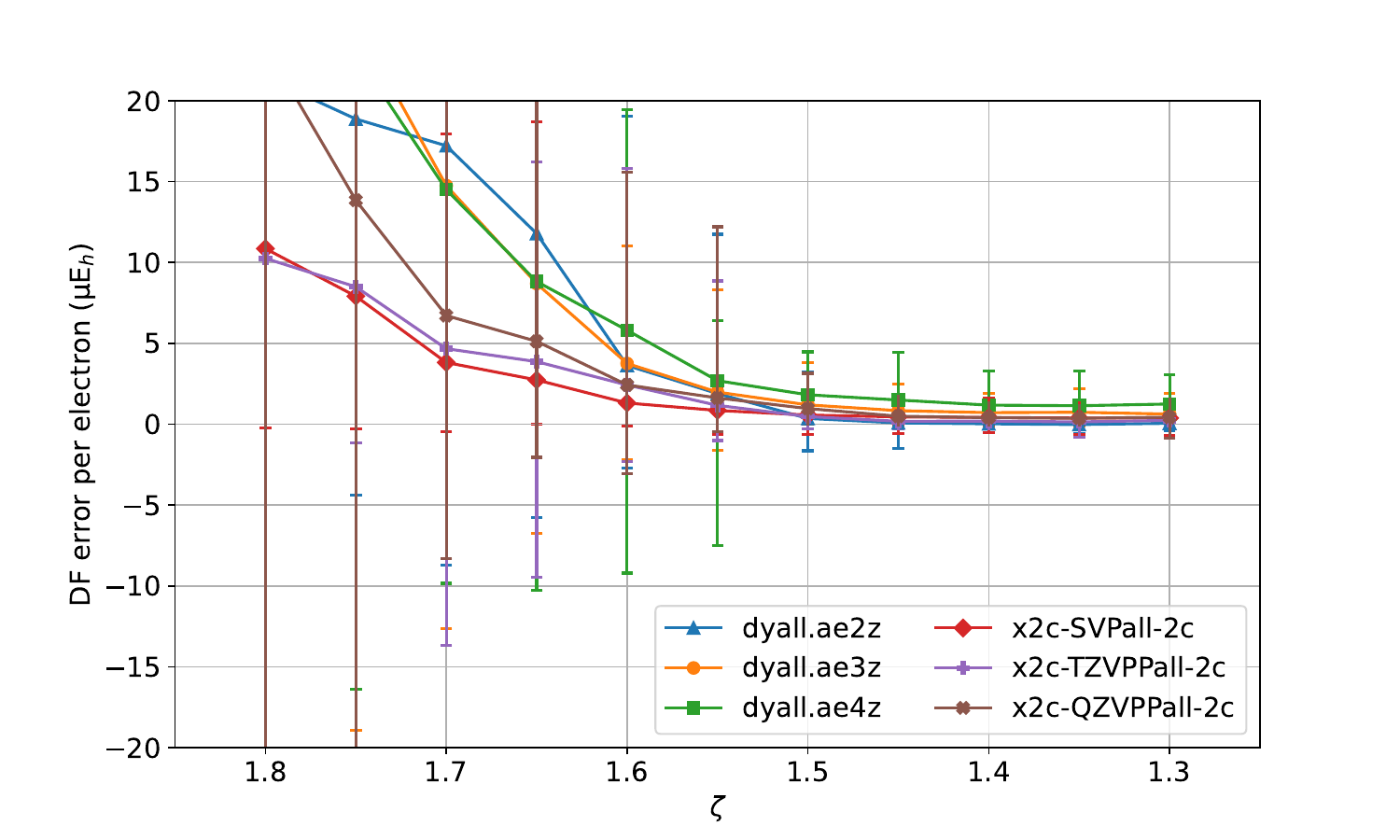}
    \caption{}
    \label{fig:rel-zetas-mp2}
\end{subfigure}
\caption{Variation of DF errors of (relativistic) (a) X2C-HF and (b) X2C-MP2 energies of the TS2 training set vs the $\zeta$ model parameter of the \name generator.}
\end{figure}

Parameters $\tau_{1,2}$ and $\zeta$ were trained again using the relativistic HF and MP2 energies of the TS2 set.
$\tau_2$ was varied first, with $\tau_1=10^{-8}$ and $ \zeta=1.3$. Here we start with $\zeta=1.3$ instead of $\zeta=1.2$ like in the non-relativistic case because $\zeta=1.2$ is too small and causes numerical issues with the relativistic OBSs.
(\cref{fig:rel-tau2-hf,fig:rel-tau2-mp2}); just as in nonrelativistic computations $\tau_2=10^{-5}$ ensures sufficient convergence of the DF errors.
$\tau_1$ was then varied with fixed $\tau_2=10^{-5}$ and $\zeta = 1.3$ (\cref{fig:rel-tau1-hf,fig:rel-tau1-mp2}). In contrast to the nonrelativistic case, a much smaller $\tau_1$ is needed for sufficient convergence of the DF errors.
The variation of relativistic DF errors with $\zeta$ (\cref{fig:rel-zetas-hf,fig:rel-zetas-mp2}) is similar to the nonrelativistic case, with $\zeta = 1.4$ deemed sufficient. Yet again, this is a much smaller ratio than the ratios used by even-tempered DFBS generators.\cite{KAS:stoychev:2017:JCTC,pu2025enhancingpyscfbasedquantumchemistry}

The optimal values of the parameters \name are briefly tabulated in \cref{tab:madf-params}.

\subsection{Assessment of DFBSs generated by the \name generator.}

\subsubsection{The G2 set}\label{sec:result-g2}
\begin{figure}[htp]
\begin{subfigure}{\textwidth}
    \centering
    \includegraphics[width=0.8\textwidth]{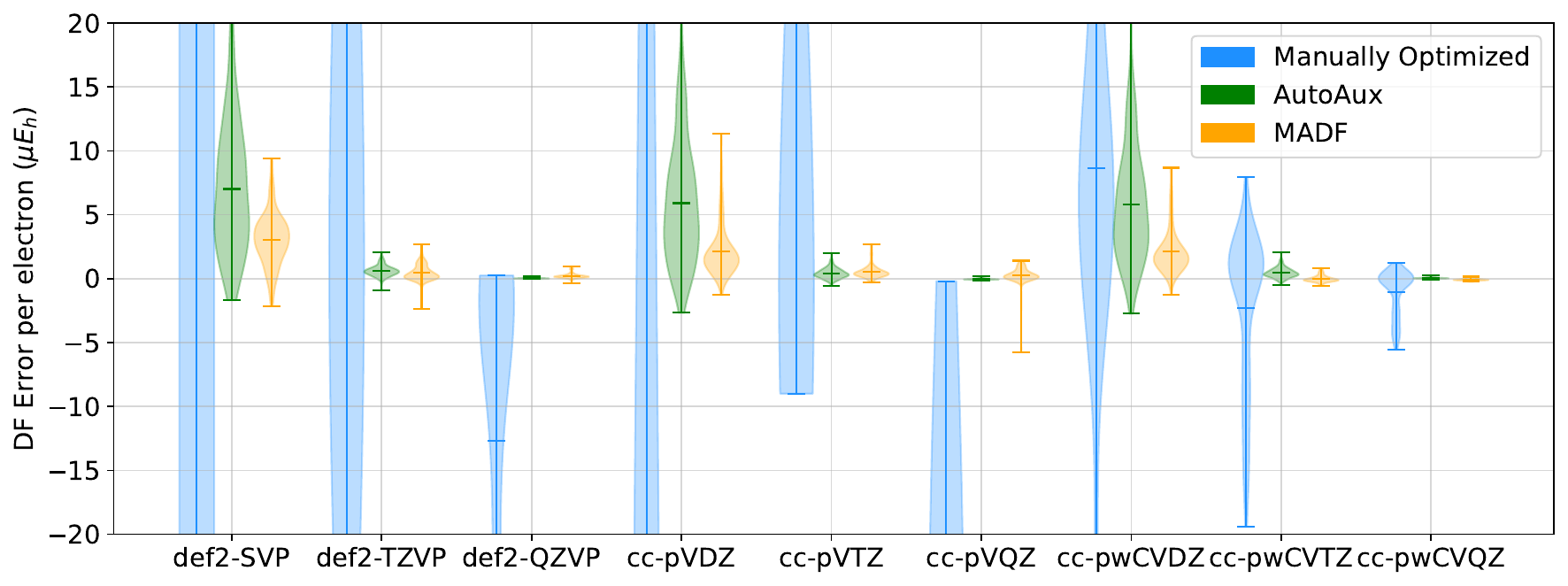}
    \caption{}
    \label{fig:g2-set-hf}
\end{subfigure}
\bigskip
\begin{subfigure}{\textwidth}
    \centering
    \includegraphics[width=0.8\textwidth]{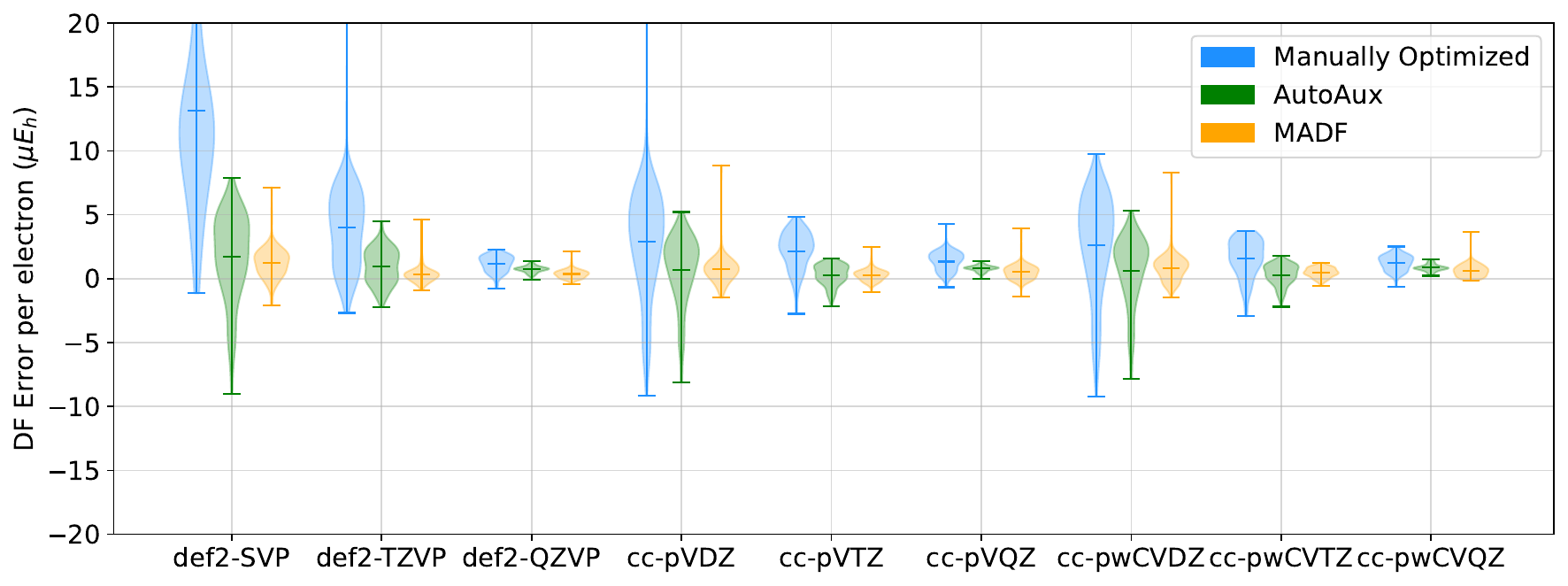}
    \caption{}
    \label{fig:g2-set-mp2}
\end{subfigure}
\begin{subfigure}{\textwidth}
    \centering
    \includegraphics[width=0.8\textwidth]{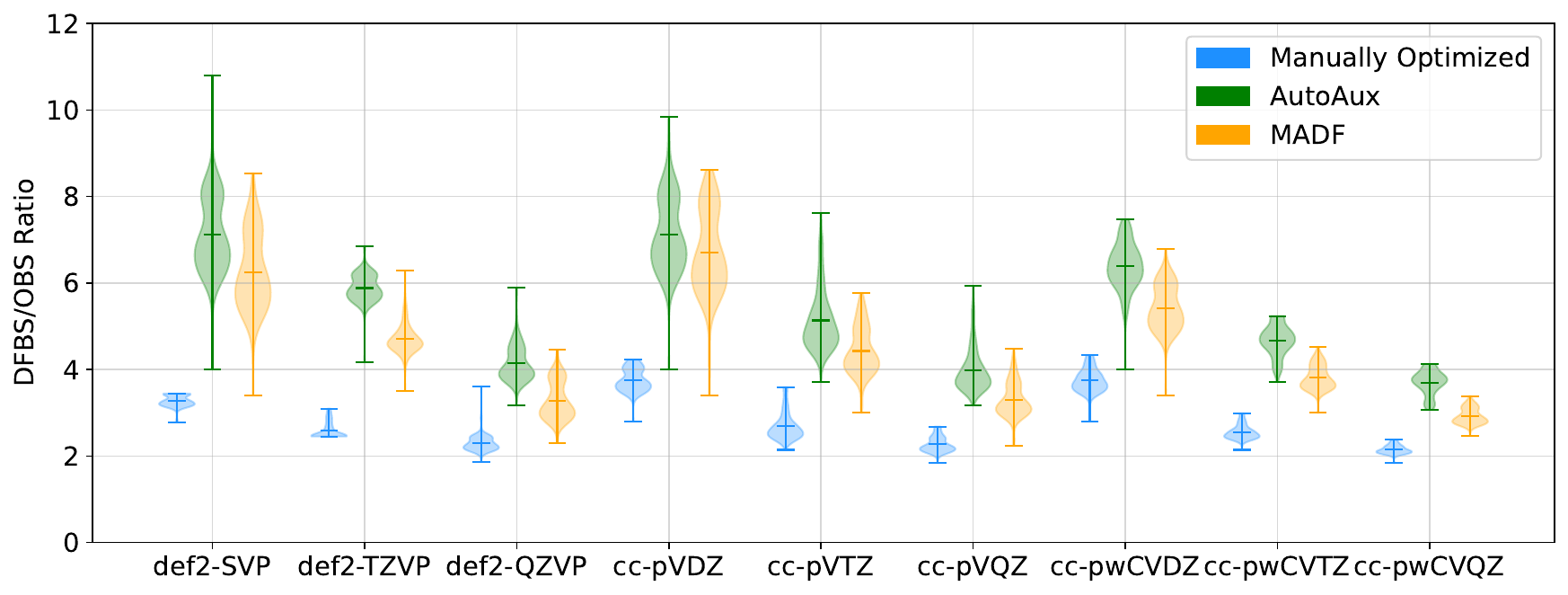}
    \caption{}
    \label{fig:g2-set-ratios}
\end{subfigure}
\caption{Comparison of DF errors of (a) HF and (b) MP2 energies of the G2 test set obtained with manually-optimized, AutoAux, and \name DFBSs. (c) Comparison of DFBS sizes, relative to that of the corresponding OBS.}
\end{figure}

The DF errors of HF and MP2 energies of the G2 test set produced with DFBS generated by the \name generator with the recommended model parameters were compared against those with manually-optimized DFBS as well as against those generated by the AutoAux generator. The \name DFBSs produce smaller DF errors in HF energies than the manually-optimized DFBS (\cref{fig:g2-set-hf}), since manually-optimized DFBSs are solely optimized to reduce DF-/RI-MP2 errors. However, the \name DFBSs also outperform the manually-optimized counterparts for MP2 energies (\cref{fig:g2-set-mp2}). However, the manually-optimized DFBS are significantly more compact (by as much as a factor of 2) than the \name counterparts, although the gap decreases significantly for quadruple-zeta basis sets (\cref{fig:g2-set-ratios}). The comparison to the AutoAux generator is more interesting. Despite fewer parameters, \name DFBSs produce similar DF errors in HF and MP2 energies.
This is especially striking since the \name DFBS are more compact than the AutoAux counterparts, sometimes by a significant margin (such as for def2-TZVP).

\begin{figure}[htp]
\begin{subfigure}{\textwidth}
    \centering
    \includegraphics[width=0.5\textwidth]{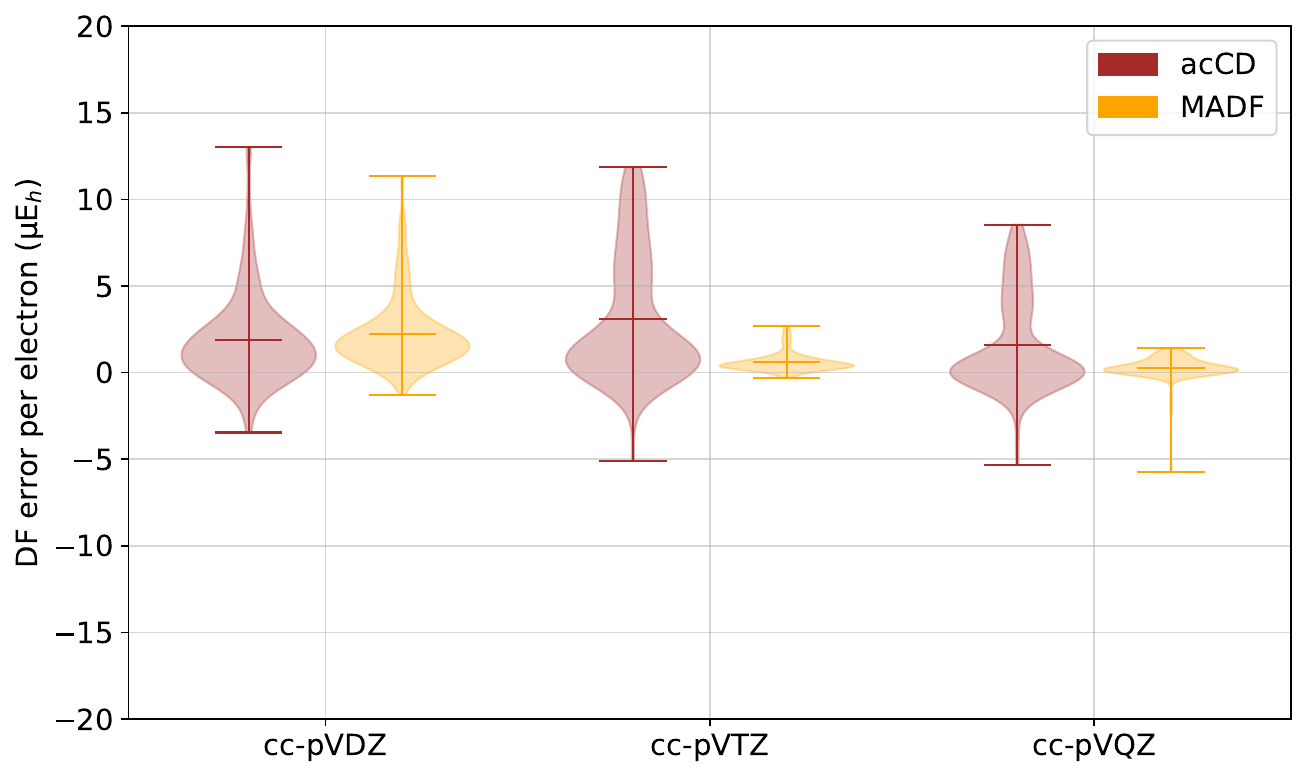}
    \caption{}
    \label{fig:g2-set-accd-hf}
\end{subfigure}
\begin{subfigure}{\textwidth}
    \centering
    \includegraphics[width=0.5\textwidth]{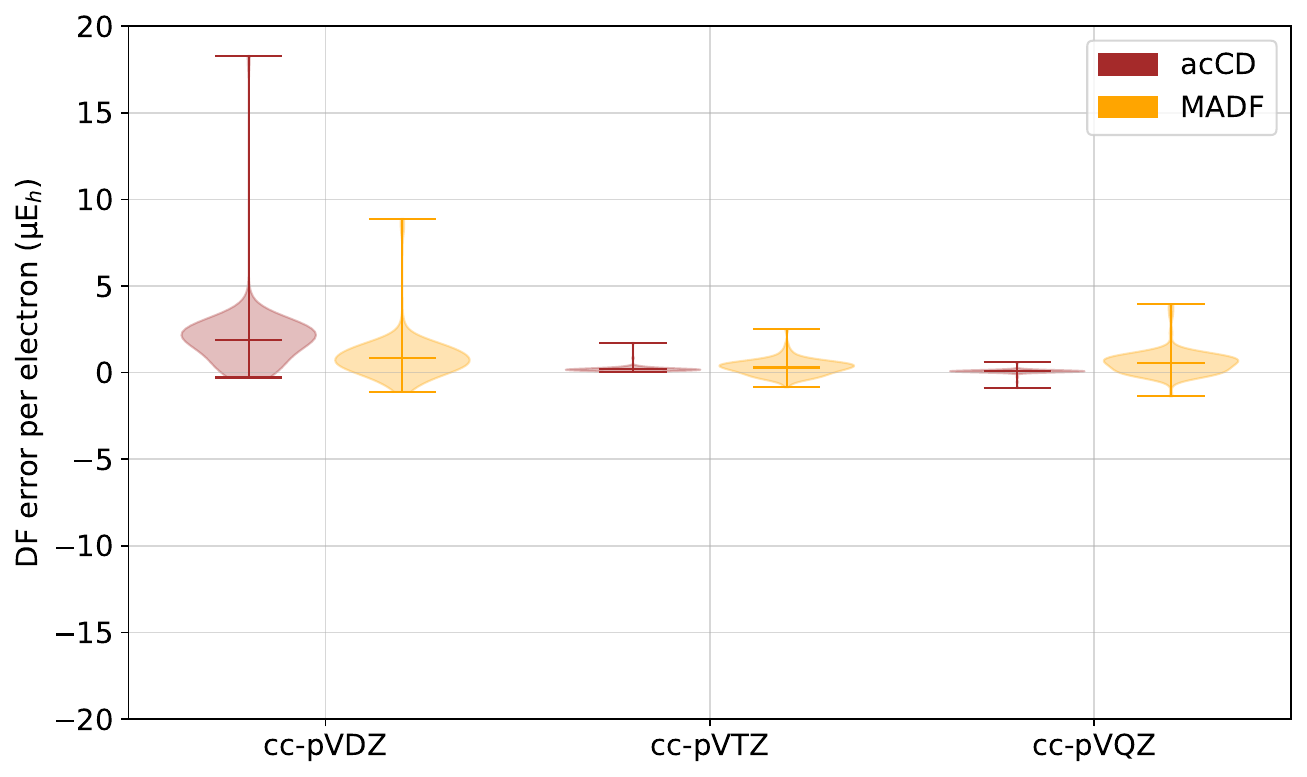}
    \caption{}
    \label{fig:g2-set-accd-mp2}
\end{subfigure}
\begin{subfigure}{\textwidth}
    \centering
    \includegraphics[width=0.5\textwidth]{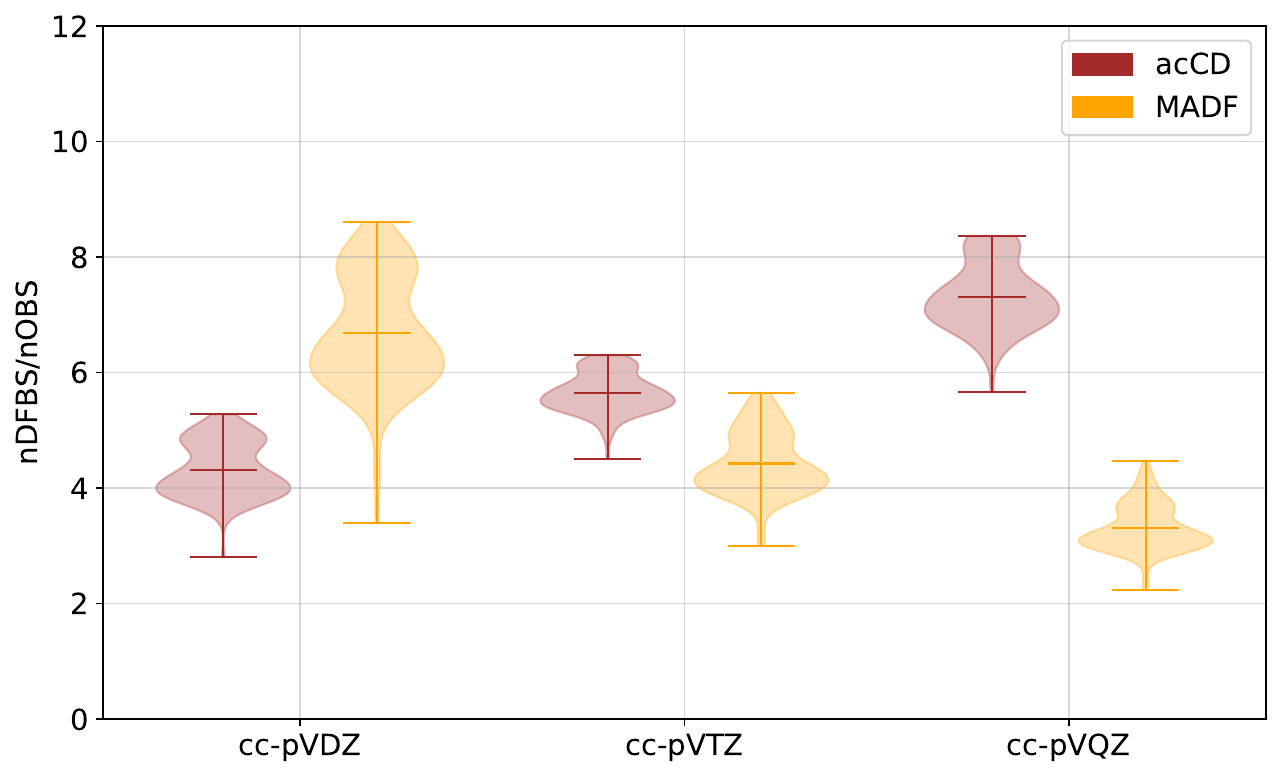}
    \caption{}
    \label{fig:g2-set-accd-ratios}
\end{subfigure}
\caption{Comparison of DF errors of (a) HF and (b) MP2 energies of the closed-shell subset of the G2 test set obtained with acCD and \name DFBSs. (c) Comparison of DFBS sizes, relative to that of the corresponding OBS. For acCD, the number of contracted DF AOs was used.}
\end{figure}

We have also compared the performance of primitive DFBSs produced by \name with contracted DFBSs produced by acCD. From \cref{fig:g2-set-accd-hf} it is clear that \name DFBSs produce smaller DF errors in HF energies compared to acCD DFBSs. As can be seen in \cref{fig:g2-set-accd-mp2} the MP2 errors for cc-pVDZ are smaller with MADF compared to acCD and for cc-pVTZ and cc-pVQZ the magnitude MP2 errors are $< 5 \mu E_h$ per electron, which is sufficient. MADF also provides significantly smaller DFBSs than that of acCD with virtually similar errors for cc-pVTZ and cc-pVQZ as seen in \cref{fig:g2-set-accd-ratios}.

\subsubsection{Ln54 and Tm60 sets}\label{sec:result-ln54-Tm60}

\begin{figure}[htp]
\begin{subfigure}{\textwidth}
    \centering
    \includegraphics[width=0.8\textwidth]{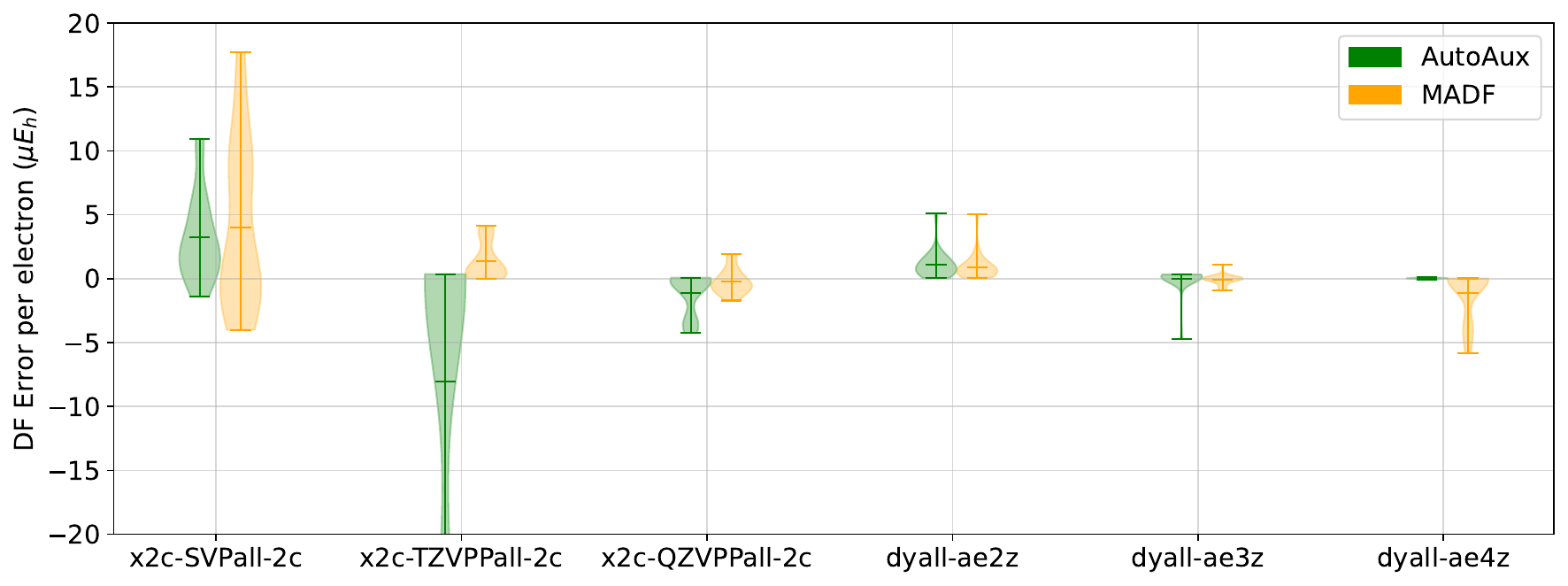}
    \caption{}
    \label{fig:Tm60-set-hf}
\end{subfigure}
\begin{subfigure}{\textwidth}
    \centering
    \includegraphics[width=0.8\textwidth]{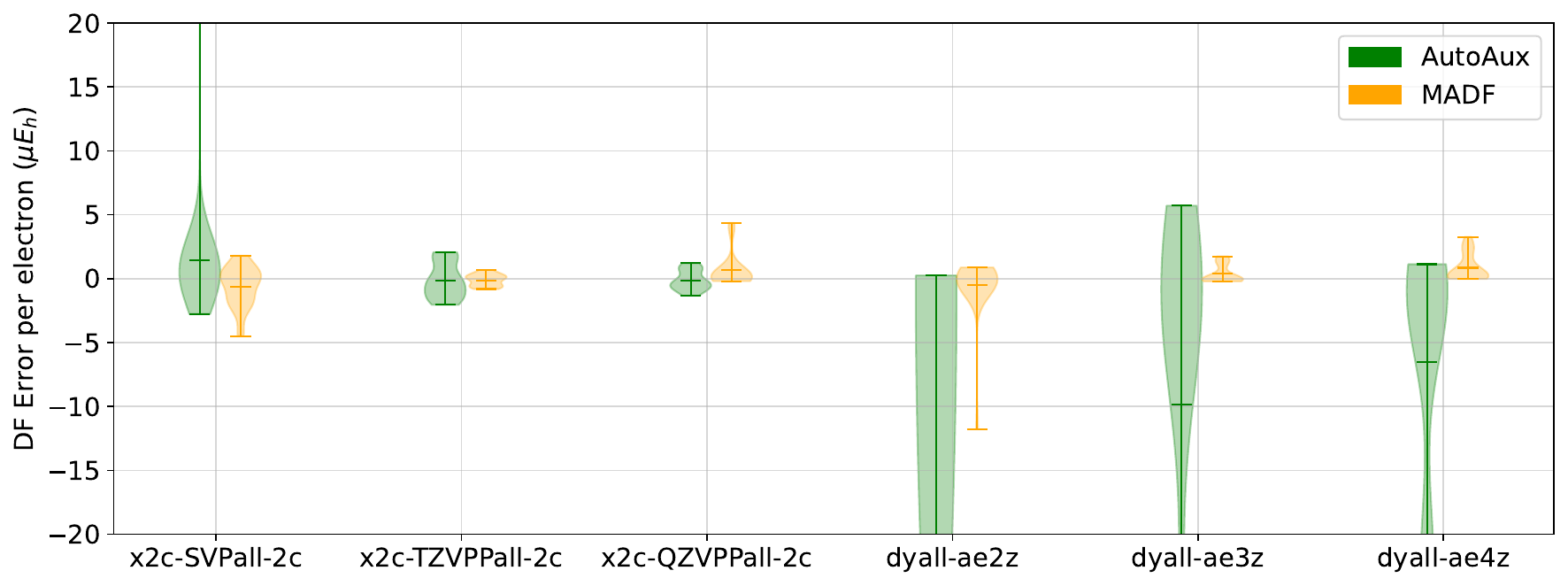}
    \caption{}
    \label{fig:Tm60-set-mp2}
\end{subfigure}
\begin{subfigure}{\textwidth}
    \centering
    \includegraphics[width=0.8\textwidth]{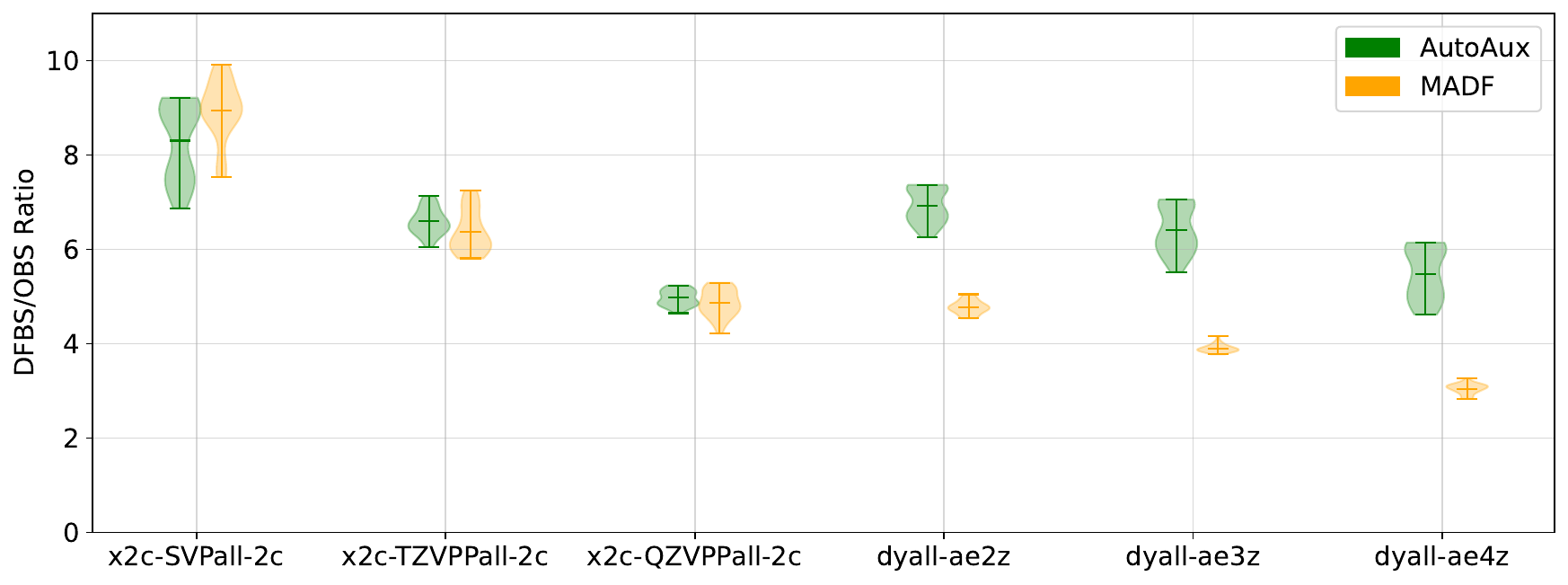}
    \caption{}
    \label{fig:Tm60-set-ratio}
\end{subfigure}
\caption{Comparison of DF errors of (a) X2C-HF and (b) X2C-MP2 energies of the Tm60 test set obtained with AutoAux and \name DFBSs. (c) Comparison of DFBS sizes, relative to that of the corresponding OBS.}
\label{fig:Tm60}
\end{figure}

\begin{figure}[htp]
\begin{subfigure}{\textwidth}
    \centering
    \includegraphics[width=0.8\textwidth]{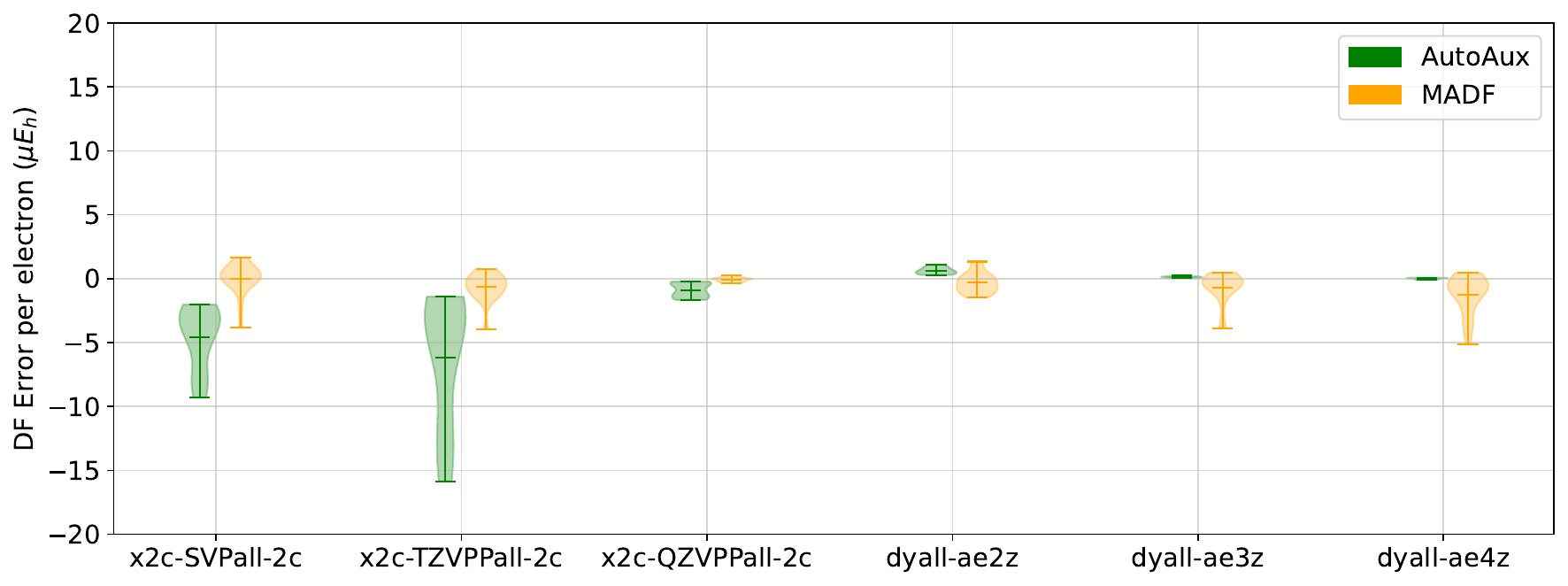}
    \caption{}
    \label{fig:ln54-set-hf}
\end{subfigure}
\begin{subfigure}{\textwidth}
    \centering
    \includegraphics[width=0.8\textwidth]{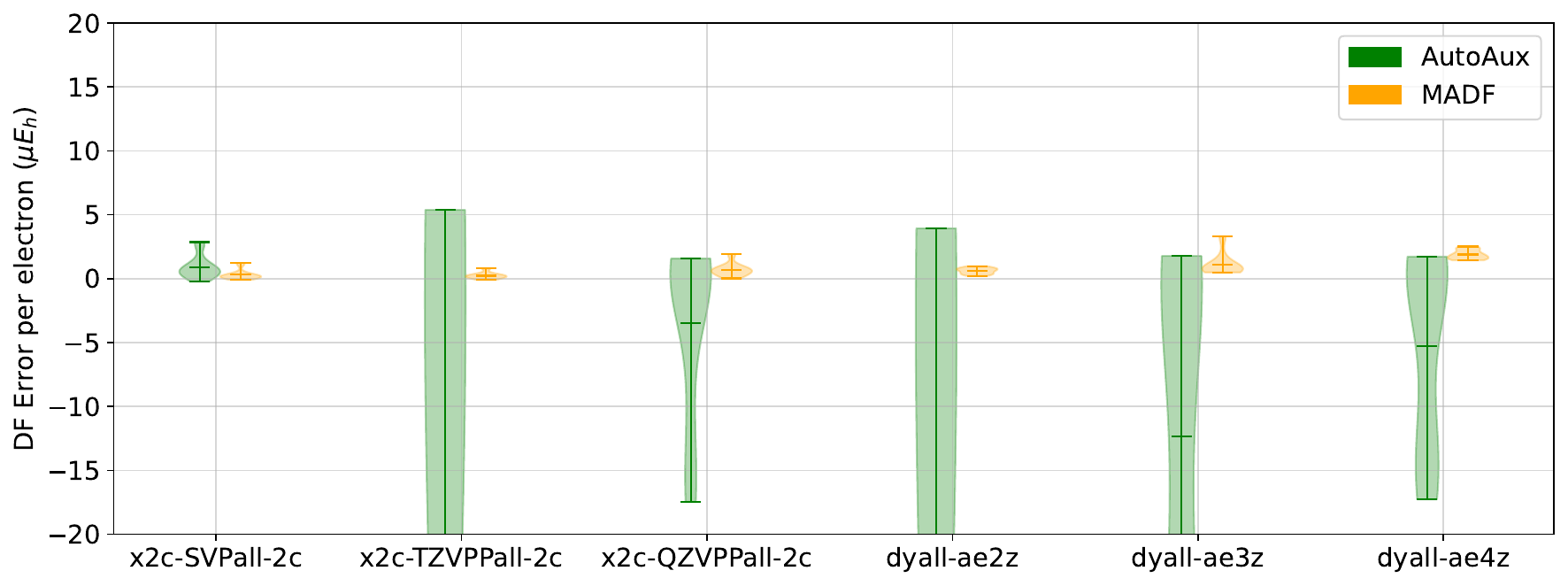}
    \caption{}
    \label{fig:ln54-set-mp2}
\end{subfigure}
\begin{subfigure}{\textwidth}
    \centering
    \includegraphics[width=0.8\textwidth]{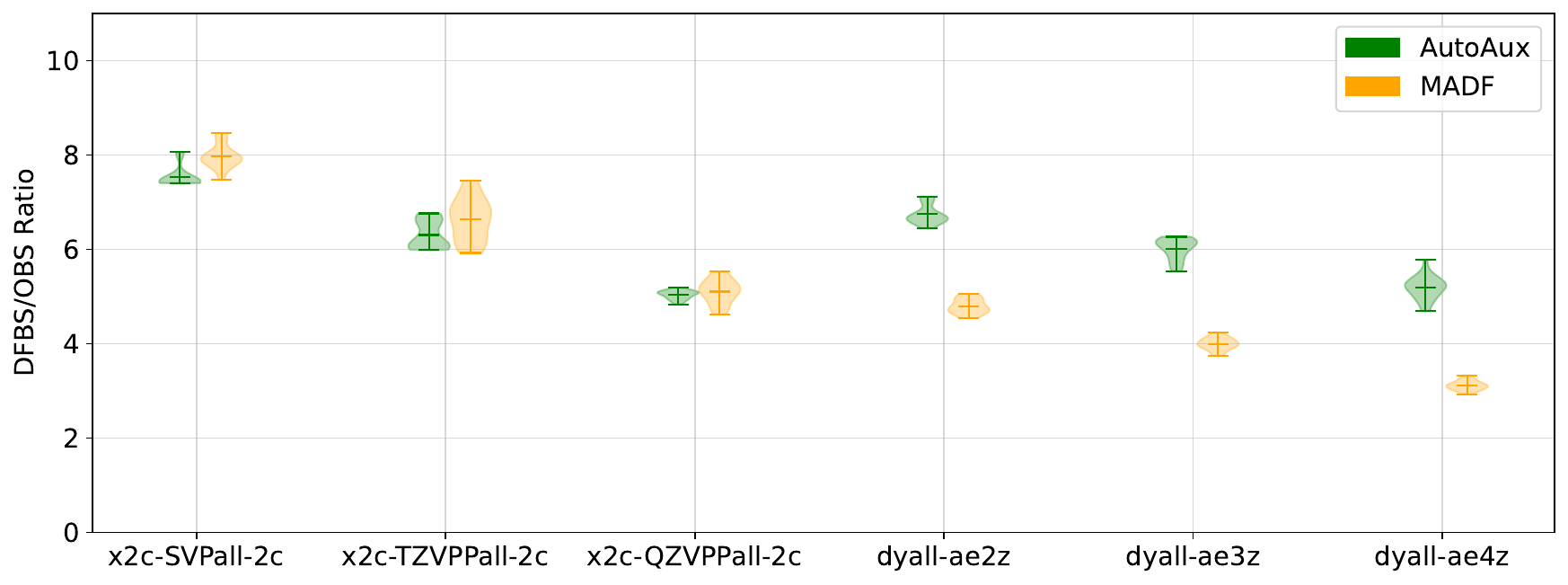}
    \caption{}
    \label{fig:ln54-set-ratio}
\end{subfigure}
\caption{Comparison of DF errors of (a) X2C-HF and (b) X2C-MP2 energies of the Ln54 test set obtained with AutoAux and \name DFBSs. (c) Comparison of DFBS sizes, relative to that of the corresponding OBS.}
\label{fig:ln54}
\end{figure}

The value of DFBS generators is especially pronounced for heavier elements where manually-optimized DFBS are simply not available.
\cref{fig:Tm60,fig:ln54} illustrate the relative performance of \name and AutoAux DFBSs for relativistic computations on molecules with d-block and f-block elements, respectively. It is evident from \cref{fig:Tm60-set-hf,fig:ln54-set-hf} that the DF errors of the X2C-HF energies are smaller with \name DFBSs than with AutoAux DFBSs, especially for the TURBOMOLE sets. For the MP2 energies (\cref{fig:Tm60-set-mp2,fig:ln54-set-mp2})  \name DFBSs are also more accurate, especially for the Dyall basis sets. Although for the TURBOMOLE OBSs \name DFBS are slightly larger than the AutoAux counterparts, the smaller errors (especially for the triple-zeta OBS) make the increase in the basis set palatable. For the Dyall basis sets \name generated significantly smaller DFBSs than AutoAux, while producing comparable DF errors of HF energies and much smaller DF errors of MP2 energies.

\section{Summary}\label{sec:summary}

This work introduced the \name algorithm for generating density-fitting basis set composed of primitive solid-harmonic Gaussian AOs and suited for nonrelativistic and relativistic (all-electron) computations with mean-field and correlated electronic structure models. Not using even-tempered AO sets for spanning the AO product space, as in other comparable generators,\cite{KAS:stoychev:2017:JCTC,pu2025enhancingpyscfbasedquantumchemistry} produces compact and accurate candidate pools of DFBS AOs while sufficiently reducing the numerical redundancy. The subsequent pruning of the candidate DFBS pool based on the 2-body energy as the importance metric allows to naturally reduce the relevant range of angular momentum channels and exponent spans with minimal use of heuristics. Generation of primitive (rather than contracted) DF AOs allows to keep the integral evaluation costs optimally low. Using only 3 model parameters \name generates basis sets that match or exceed the accuracy-to-size ratio of the state-of-the-art DFBS generators; the comparison is particularly favorable with OBSs suited for relativistic all-electron computations.
Computations utilizing \name DFBS, of course, can also benefit from system-specific DFBS compressions.\cite{KAS:kallay:2014:JCP,VRG:schurkus:2017:JCP}

Although picture change and other relativistic effects on the 2-particle interaction were not considered here, it will be interesting to examine the impact of such effects on the optimal \name model parameters in the future; it is known that there are nontrivial differences in the requirements on the DFBS that stem from relativistic contributions to the effective 2-particle interactions.\cite{KAS:kelley:2013:JCP} 
Even in the nonrelativistic framework much more testing of the approach will be clearly needed. Most important is to test the performance in the context of higher-order correlated methods (especially coupled cluster), for DF approximation of non-Coulomb operators, for properties other than energy, and for other basis set families (e.g., multiply-augmented correlation-consistent sets, ANO, etc.).
Early experiences with \name DFBSs appear promising. For example, we recently utilized the \name DFBS generator for explicitly-correlated (F12) coupled-cluster computations with aug-cc-pV7Z OBS (for which no manually-optimized DFBS is available) where we found that the DF errors with the \name DFBS are significantly smaller than the use of the manually-optimized aug-cc-pV6Z-RIFIT basis.\cite{VRG:powell:2025:JCTC} These and other experiences with \name DFBSs will be reported elsewhere.

\begin{suppinfo}

\name density-fitting basis sets generated in this work in the Gaussian 94 basis set format; molecular geometries of all species.

\end{suppinfo}

\begin{acknowledgement}
Work by KAS was supported by the U.S. Department of Energy, Office of Science, Office of Advanced Scientific Computing Research, and Office of Basic Energy Sciences, Scientific Discovery through the Advanced Computing (SciDAC) program under Award DE-SC0022263. Work by EFV was supported by the US Department of Energy, Office of Science, via award DE-SC0022327. This project used resources of the National Energy Research Scientific Computing Center, a DOE Office of Science User Facility supported by the Office of Science of the U.S. Department of Energy under Contract DE-AC02-05CH11231 using NERSC award ERCAP-0024336. The authors acknowledge Advanced Research Computing at Virginia Tech (https://arc.vt.edu/) for providing computational resources and technical support that have contributed to the results reported within this paper. The development of the \code{Libint}  software library is supported by the Office of Advanced Cyberinfrastructure, National Science Foundation (Award OAC-2103738).
\end{acknowledgement}

\begin{appendix}
\section{Superposition of atomic densities (SOAD)}\label{app:soad}

Unlike the SAD method,\cite{VRG:vanlenthe:2006:JCC}
with computes the Fock matrix from the atomic (optionally, spin-free) ensemble density matrix computed in OBS self-consistently, SOAD Fock matrix is computed from fixed spin-free ensemble occupancies ${\bf n}_{\mathrm{SOAD}}$ of the minimal basis set (MBS) AOs, without self-consistency.
${\bf n}_{\mathrm{SOAD}}$ for a single atom  in an {\em orthonormal} MBS is obtained by smearing each subshell's electrons evenly among all of its orbitals;
e.g., for a carbon atom occupancies of 1s, 2s, 2p$_x$, 2p$_y$, and 2p$_z$ MBS AOs are 2, 2, 2/3, 2/3, and 2/3, respectively. This recipe has been referred to as average-of-configuration (AOC)\cite{KAS:zerner:1989:IJQC} and canonical ensemble\cite{KAS:jansik:2009:PCCP}. The number of electrons in each subshell is defined by the electron configuration of that atom's neutral ground state taken from NIST Atomic Spectra Database\cite{VRG:kramida:1999:}.
For a molecule ${\bf n}_{\mathrm{SOAD}}$ is obtained by concatenation of the atomic occupancies, without taking into account the effects of nonunit molecular MBS metric. In this work the MINI basis\cite{VRG:andzelm:1984:PSDV1GBSfMC,KAS:vanduijneveldt:1971:IBM} was used as MBS for $Z\leq 54$, and the ANO-RCC-MB basis\cite{VRG:roos:2005:CPL} for $54<Z\leq 96$.
${\bf n}_{\mathrm{SOAD}}$ is used to construct SOAD Fock matrix in OBS AO basis, $\mathbf{F_{\mathrm{SOAD}}}$, as follows:
\begin{equation}\label{eq:soad-fock}
    (\mathbf{F}_{\mathrm{SOAD}})^\mu_\nu = h^\mu_\nu + \sum_{m }  ({\bf n}_{\mathrm{SOAD}})_m [ 2( \mu \nu | m m )  - ( \mu m | \nu m ) ],
\end{equation}
where $h^\mu_\nu$ are the matrix elements of the 1-body (core) Hamiltonian (sf-1eX2C\cite{KAS:filatov:2003:JCP, VRG:kutzelnigg:2005:JCP, VRG:peng:2013:JCP} in this work) , $m$ runs over MBS AOs and $\mu,\nu$ over OBS AOs. It should be noted that \cref{eq:soad-fock} is similar to the projection-free guess proposed by Norman and Jensen\cite{KAS:norman:2012:CPL} for Dirac-Hartree-Fock/Kohn-Sham.

\end{appendix}

\bibliography{vrgrefs, KAS, misc}

\end{document}